\documentclass{aa}  

\usepackage{graphicx}
\usepackage{amsmath, amssymb}
\usepackage{txfonts}
\usepackage{hyperref}
\usepackage[dvipsnames]{xcolor}
\usepackage{placeins}

\begin{document} 

   \title{Near-Eddington mass loss of hydrogen-rich Wolf-Rayet stars}
   \subtitle{Consequences of temperature, metallicity, and continuum acceleration}

   \author{
   R.R. Lefever\inst{\ref{inst:ari}} 
   \and 
   A.A.C. Sander\inst{\ref{inst:ari}} 
   \and
   M. Bernini-Peron\inst{\ref{inst:ari}} 
   \and
   G. Gonz\'alez-Tor\`a\inst{\ref{inst:ari}} 
   \and
   N.M. Moens\inst{\ref{inst:KUL}}
   \and
   F. Najarro\inst{\ref{inst:cab}}
   \and 
   E.C. Sch\"osser\inst{\ref{inst:ari}}
   \and
   G.N. Sabhahit\inst{\ref{inst:AOP}}
   \and
   J.S. Vink\inst{\ref{inst:AOP}}
   }

   \institute{%
   {Zentrum f{\"u}r Astronomie der Universit{\"a}t Heidelberg, Astronomisches Rechen-Institut, M{\"o}nchhofstr. 12-14, 69120 Heidelberg\label{inst:ari}}\\
   \email{roel.lefever@uni-heidelberg.de}
   \and
   {Instituut voor Sterrenkunde, KU Leuven,
    Celestijnenlaan 200D, 3001 Leuven, Belgium \label{inst:KUL}}     
   \and
   {Departamento de Astrof\'{\i}sica, Centro de Astrobiolog\'{\i}a, (CSIC-INTA), Ctra. Torrej\'on a Ajalvir, km 4,  28850 Torrej\'on de Ardoz, Madrid, Spain\label{inst:cab}}
   \and
   {Armagh Observatory and Planetarium, College Hill, BT61 9DG Armagh, Northern Ireland, UK\label{inst:AOP}}
  }
   \date{Received 19 May 2025 / Accepted 6 July 2025}
 
  \abstract
   {Very massive clusters and regions of intense star formation such as the center of our Milky Way contain young, hydrogen-burning stars very close to the Eddington Limit. 
   Formally classified as hydrogen-rich Wolf-Rayet stars, the winds and spectra of these stars are distinctively different to the more evolved, classical Wolf-Rayet (cWR) stars.}
   {In this work, we focus on the so far less examined wind regime of later-type WNh stars, which have evolved away from the zero age main sequence. 
   Our aim is to uncover the wind physics in this regime and determine similarities and differences to other wind regimes.}
   {We create sequences of hydrodynamically-consistent atmosphere models resembling massive, slightly evolved WNh stars very close to the Eddington Limit. 
   Our models span temperatures between 21 and 45\,kK and metallicities between 1.2 and 0.02 solar. 
   We also use the opportunity to predict spectra in a larger metallicity range than thus far covered by resolved observations.}
   {We find an overall downward trend of the mass-loss rate with increasing temperature and decreasing metallicity. 
   However, at SMC metallicities and above, we find a maximum in the wind efficiency with the mass-loss eventually decreasing again at lower temperatures. 
   For intermediate metallicities, we also find strong discontinuities in the mass-loss trends, which do not appear at high or very low metallicities. 
   For the lowest metallicities, a more homogeneous behavior is obtained without any maximum in the wind efficiency. 
   The terminal velocities are generally higher for hotter temperatures. 
   For cooler temperatures, the combined effect of metallicity and mass-loss change significantly reduces the changes in terminal velocity with metallicity.}
   {Contrary to cWR stars, the spectral appearance of late-type WNh stars rules out supersonic winds launched at the hot iron bump. 
   Instead, a more extended quasi-hydrostatic regime is necessary.
   The proximity to the Eddington limit and the complex interactions cause a lot of substructure in the global wind parameter trends. 
   While the strong discontinuities show resemblances to the bi-stability jump predicted for the B-supergiant regime, our models reveal a more complex origin. 
   At sub-SMC metallicity, iron is no longer a major key for setting the mass-loss rate in this WNh regime. 
   Instead, other elements (e.g. nitrogen) and continuum contributions become important. 
  }

   \keywords{
   stars: Wolf-Rayet --
   stars: atmospheres --
   stars: winds, outflows --
   stars: early-type --
   stars: mass-loss
    }

   \maketitle

%-------------------------------------------------------------------

\section{Introduction}\label{sec:intro}

Wolf-Rayet (WR) stars are a subset of massive stars with particularly extreme properties. 
Found in the upper left regions of the Hertzsprung-Russell diagram (HRD), these stars typically have very high luminosities and temperatures.
As such, these stars' powerful radiation serves as a strong ionizing source for their local environments \citep[e.g.,][]{smith2002,crowther-hadfield2006, hainich2015}. 
Being also subject to strong, line-driven stellar winds with mass-loss rates on the order of $10^{-5}\ldots 10^{-4}\, M_\odot\,\mathrm{year}^{-1}$ \citep{crowther2007}, WR stars are important engines of mechanical feedback
\citep[e.g.,][]{sander-vink2020} and contribute strongly to local chemical enrichment \citep{maeder1983,dray2003,farmer2021}.
Spectroscopically, WR stars are characterized by their strong and broad emission lines, originating from their stellar winds. 
Effectively, WR stars eject their outermost layers \citep{conti1983,smith2014}. 
WR stars with sufficiently high mass-loss rates eject their dense outermost layers, obscuring their inner hydrostatic layers.
This causes observers to see the light from the irradiated wind material instead \citep{hamann1985, shenar2020}. 
Based on the presence and strength of certain lines present in their spectra, WR stars are further classified into WN stars with strong nitrogen signatures, WC stars with carbon features present, and WO stars with prevalent oxygen lines \citep{hiltner1966,smith1968}. 

With respect to their evolution status, most WR stars are evolved objects that have already finished their central hydrogen burning. 
In many cases the absence of hydrogen in their spectrum provides clear evidence that the stars have lost their outer layers and are now in the status of core-He burning. 
Others still contain hydrogen, but in a significantly depleted fraction such that their parameters also suggest them to be core-He burning. 
Such core-He burning WR stars are also called ``classical'' WR (cWR) stars and make about 90\% of the known massive WR population in the Galaxy \citep[][]{crowther2007,shenar2019}.

Aside from the classical WR stars, a minority of the WR stars have significantly higher masses and comparably high hydrogen surface abundances. 
Such stars typically have less dense winds than cWR stars and are thought to be still core-hydrogen burning \citep{dekoter1997}, thereby effectively being main sequence objects. 
Spectroscopically, these stars have the spectral type WNh. 
Yet, this designation formally just implies a WN subtype with the presence of hydrogen. 
Evolved WR stars with leftover hydrogen at the surface can also have a WNh spectral type. 
Therefore, a quantitative spectral analysis is usually required to constrain the evolution status of a WNh star.

Eventually, most WR stars are massive enough to collapse into stellar-mass black holes \citep{woosley2020,higgins2021,vink2021}. 
As the last longer-lived stage before core collapse, WRs mark an important anchor to observationally constrain massive star evolution and the resulting black hole masses, in particular given the recent insights from gravitational wave events \citep[e.g.,][]{ligo2021}. 

Despite the important role of WR stars, the complexity of their atmospheres and winds has long prevented deeper physical insights. 
While traditional quantitative spectroscopic analysis has been done for a considerable sample of WR stars in different galaxies \citep[e.g.,][]{hillier1999,Crowther2002,Hamann2006,Liermann2010,Sander2014,Tramper2015,Shenar2016}, many open questions about the underlying wind physics remain. 
In many cases, the emission-line dominated spectra prevent robust conclusions on the stellar masses and thus empirical laws between the derived mass-loss rates and basic stellar parameters have inherent uncertainties. 
Moreover, the wind dynamics in such studies are almost always approximated by a $\beta$-velocity law, which has been proven to be too simple for many cWR stars \citep{graefener2005,sander2020driving,poniatowski2021}
% https://www.aanda.org/articles/aa/abs/2021/03/aa39595-20/aa39595-20.html
and limits the spectral diagnostics, in particular when relying only on optical spectra \citep[e.g.][]{lefever2023}. 
The complex wind structure has also been confirmed by multi-D, radiation-hydrodynamic simulations \citep{moens2022}. 
Dynamically-consistent atmosphere calculations \citep{graefener2005,sander2017hydro,sundqvist2019} are able to overcome the typical assumption of a prescribed velocity structure, usually in the form of a $\beta$-law, by solving the hydrodynamic equation of motion consistently with a comoving-frame calculation of the radiative transfer \citep{graefener2005,sander2017hydro,sundqvist2019}. 
With this approach, the wind parameters can be inherently linked to the stellar parameters, thereby enabling a new approach to decipher the wind physics of WR stars.

Recently, this method has been very successful in establishing the first theoretical prescription for the mass-loss rates of cWR stars \citep{sander2020driving,sander-vink2020,sander2023}. 
\citet{graefener2008} applied an earlier version of this technique to the regime of WNh stars, based on an analysis of the protoype WR\,22. 
Yet, their derived mass of $78.1\,M_{\odot}$ for the WR star is much higher than the recent orbital analysis by \citet{Lenoir-Craig2022} yielding $\sim$$56\,M_\odot$. 
In addition, the computational capabilities at that time severely limited the amount of elements and ions to be considered in the model calculations. 
All these aspects underline the need for revisiting this regime with new, improved calculations.

In this work, we therefore address the regime of WNh stars with a modernized dynamically consistent approach and a significantly extended set of atomic data compared to \citet{graefener2008}, shown in Table \ref{tab:abundances}. 
Focusing in particular on the later types, we find notable differences and sudden regime changes with respect to the wind launching and the resulting mass-loss rates.

We will outline the code framework and model sample in Sect.\,\ref{sec:methods}. 
Afterwards, the findings from the model sequence are presented in Sect.\,\ref{sec:results}. 
%Additionally, the effect of lower-than-solar metallicities in this temperature range are also touched on in this section. 
%Finally, the imprint on the model spectra of this jump is also detailed. 
The physical implications of our results will be discussed in Sect.\,\ref{sec:discussion}, along with a comparison to observed WNh stars. In Sect.\,\ref{sec:conclusions}, we draw our conclusions and give an outlook on the implications for further studies.

%--------------------------------------------------------------------
\section{Methods}\label{sec:methods}

To model the atmospheres of WNh stars, we employ the Potsdam Wolf-Rayet (PoWR) code \citep[][]{graefener2002, hamann2003, sander2015}. 
Assuming a spherically symmetric star with a stationary outflow,  PoWR computes the non-LTE population numbers together with a radiative transfer solution in the co-moving frame.
Input parameters for our models are the stellar mass $M_\ast$, luminosity $L$, temperature $T_\ast$ \, and the chemical abundances (typically in mass fractions) $X_i$ of the desired elements.
The temperature $T_\ast$ marks an effective temperature corresponding to the inner boundary radius $R_\ast$ defined at a specific Rosseland continuum optical depth $\tau_\text{cont,max}$. 
The $L$, $T_\ast$ and $R_\ast$ parameters are related by the Stefan-Boltzmann relation $L_\ast = 4\pi R_\ast^2 \sigma_\mathrm{SB} T_\ast^4$.
Following the calculation approach from \citet{Sabhahit2023}, we choose $\tau_\text{cont,max} = 5$. 
This is fully sufficient for the prediction of the WNh-spectra and the determination of the wind parameters. 
This approach avoids an explicit modeling of the deeper atmosphere-envelope connection, which will be explored in a separate project \citep{Josiek+2025}. 
Furthermore, as is shown in Sec. \ref{subsec:comparison}, we find fundamental differences in our models compared to classical wind launching where the wind is initiated at the ``hot iron bump'', a large increase in opacity due to the excitation of M-shell Fe ions (\ion{Fe}{ix} to \ion{Fe}{XVI}), typically at temperatures of  $\sim$200 kK.
In the models used in this work, the resulting inner boundary velocity is typically between $0.01$ and $0.1\,\mathrm{km\,s^{-1}}$ for models with lower mass-loss rates (i.e., higher temperature, lower metallicity, or both) but can reach up to $1.2\,\mathrm{km\,s^{-1}}$ for the models with the highest mass loss (low $T_\ast$ and $Z \geq Z_\odot$). 
For this regime, we are reaching the limit of what is feasible with this methodology as we will discuss later.

For our study, we specifically use the PoWR$^\textsc{hd}$ branch \citep{sander2017hydro,sander-vink2020,sander2023} which solves the hydrodynamic equation of motion to consistently predict the wind velocity and density rather than assuming a pre-specified $\beta$-velocity law. 
Accounting for gravity, radiation pressure and gas pressure, the 1D stationary equation of motion reads

\begin{equation}\label{eq:hydro-original}
    \varv \frac{\mathrm{d}\varv}{\mathrm{d} r} + \frac{GM}{r^2} = a_\mathrm{rad}(r) + a_\mathrm{press}(r)\text{.}
\end{equation}

To obtain the wind velocity structure $\varv(r)$, Eq.\,\eqref{eq:hydro-original} can be rewritten as

\begin{equation}\label{eq:hydro-solving}
    \frac{\mathrm{d} \varv}{\mathrm{d} r} = -\frac{g}{\varv} \frac{\tilde{\mathcal{F}}}{\tilde{\mathcal{G}}},
\end{equation}

\noindent where, with the definition $\Gamma_\mathrm{rad}(r) := a_\mathrm{rad}(r)\, /\, g(r)$, 

\begin{equation*}
\begin{split}
    \tilde{\mathcal{F}} & := 1 - \Gamma_\mathrm{rad}(r) - 2\frac{a(r)^2\,r}{GM} + \frac{r^2}{GM}\frac{\mathrm{d}a(r)^2}{\mathrm{d}r}\,\text{, and}\\
    \tilde{\mathcal{G}} & := 1 - \frac{a(r)^2}{\varv(r)^2}.
\end{split}
\end{equation*}

\noindent Here, $a(r)$ denotes the (isothermal) sound speed, optionally adjusted for a turbulent velocity. A detailed discussion and derivation of the equations presented above is given in \citet{sander2015, sander2017hydro}. 
From a critical radius $R_\text{crit}$ defined by $\tilde{\mathcal{G}} = 0$, Eq.\,\eqref{eq:hydro-solving} is integrated inwards and outwards to obtain $\varv(r)$ (along with $\varv_\infty$). 
This integration is iterated with different mass-loss rates $\dot{M}$ until a self-consistent solution that preserves the chosen optical depth at the lower boundary is obtained \citep[see][for details]{sander2017hydro}.
Alternatively, the mass-loss rate $\dot{M}$ can also be a fixed as a input parameter while the stellar mass is adjusted instead \citep{sander2020driving}.

While previous WNh PoWR$^\textsc{hd}$ model efforts have focused on objects with effective temperatures around $45\,$kK, such as the very massive stars in NGC3603 or R136 in the LMC \citep[][]{Sabhahit2023,Sabhahit+2025}, our models in this work are based on a cooler WNh-regime, such as the stars found in the Galactic Center. 
We base our models on parameters motivated by the WN9h stars like WR\,102hb, WR\,102ea, or WR\,102d found in the Quintuplet cluster \citep[][]{Liermann2009,Liermann2010,Clark2018} and the so-called ``Peony star'' WR\,102ka \citep[WN10h][]{barniske2008,oskinova2013}. 
The late-type WNh stars show luminosities $\log L/L_\odot > 6.0$, hydrogen mass fractions $X_\text{H} < 0.5$ and $24\,\mathrm{kK} \lesssim  T_\ast \lesssim 36\,\mathrm{kK}$. 
Some of the objects in the Galactic Center \citep[e.g., LHO\,110 in][]{Liermann2010} are very similar, but their spectral appearance does not formally classify them as WNh or Ofpe/WN type. 
This also happens in our models, where in particular low-metallicity models might not formally qualify as WR, which we briefly illustrate and discuss in Sect.\,\ref{sec:results}.
Exploring the winds of the putatively H-burning late-type WNh stars and their lower-metallicity analogs, irrespective of their formal spectral classification, will be the focus of this paper.

For our study, we generated a large set of PoWR$^\textsc{hd}$ models with fixed ``typical'' late-type WNh parameters of $\log L = 6.3\, [L_\odot]$, $M_\ast = 69\, M_\odot$, and a hydrogen mass fraction of $X_\text{H} = 0.2$. 
All elements and ions with their respective chemical abundances (for metallicity $Z = Z_\odot$) are given in Table\,\ref{tab:abundances}. 
Based on these model parameters, we compute a series of models -- in total 111 -- where the temperature $T_\ast$ and metallicity $Z$ have been varied. 
The temperature ranges for each $Z$ are listed in Table\,\ref{tab:model-params}. 
We will use these temperature sequences of models to dissect the influence of $T_\ast$ on the stellar winds of late-type WNhs.
For consistent comparison with the model series in \citet{sander2023}, we also include a small, radially constant turbulent pressure of $\varv_\text{turb} = 21\,\mathrm{km\,s}^{-1}$ in the momentum equation for all our models. 
Compared to $\varv_\text{turb} = 0\,\mathrm{km\,s}^{-1}$, the value we use will not display significant differences in general behavior of the obtained wind solutions. 
Clumping is treated in the standard ``microclumping'' approximation \citep{hamann1998} with a maximum clumping factor $D = 10$ and a depth-dependent, exponential increase following the description of the so-called ``Hillier law'' \citep{Hillier2003} with a characteristic onset velocity of $\varv_\text{cl} = 100\,\mathrm{km\,s}^{-1}$.

\begin{table}[h]
    \caption{Temperature $T_\ast$ and metallicity $Z$ sequence of the models used in this study.}
    \label{tab:model-params}
    \centering
    \begin{tabular}{c|c}
         $Z\, [Z_\odot]$ & $T_\ast^a$ [kK] \\
         \hline
         1.2 & $24\,-\,31$ ; $37\,-\,39$ ; $41$ ; $43\,-\,44$ \\
         1   & $24\,-\,44$ \\
         0.8 & $25\,-\,41$ \\
         0.6 & $24\,-\,43$ \\
         0.4 & $24\,-\,42$ \\
         0.2 & $24\,-\,33$; $35$\\
         0.1 & $25\,-\,32$\\
         0.06 & $22\,-\,35$\\
         0.02 & $22\,-\,34$\\
    \end{tabular}
        \tablefoot{
        \tablefoottext{a}{$\Delta\, T_\ast = 1$ kK for continuous temperature ranges}
        % \tablefoottext{b}{See Eq.\,\eqref{eq:transformed_massloss}.}
    }
\end{table}

\section{Results}\label{sec:results}

\subsection{Global wind parameters}\label{subsec:global}

We begin with examining the temperature sequence at solar metallicity. 
Aside from the global wind parameters of the terminal wind velocity $\varv_\infty$ and the mass-loss rate $\dot{M}$, we also inspect the wind efficiency $\eta = \dot{M}\,\varv_\infty \left(L_\ast / c\right)^{-1}$ (with $c$ the speed of light in vacuum and $L_\ast$ the stellar luminosity) and the transformed mass-loss rate

\begin{equation}\label{eq:mdott}
    \dot{M}_\mathrm{t} = \dot{M}\,\sqrt{D} \left(\frac{1000\,\mathrm{km}\,\mathrm{s}^{-1}}{\varv_\infty}\right)\,\left(\frac{10^6\, L_\odot}{L_\ast}\right)^{3/4}, 
\end{equation}

\noindent as defined by \citet{graefener2013}. 
In Eq.\,\eqref{eq:mdott}, $D$ is the clumping factor \citep[see][]{hamann1998} and -- in the case of a void interclump medium and optically thin clumps -- is  the inverse of the volume filling factor, i.e. $D = f_\mathrm{V}^{-1}$. 
In our calculations for $\dot{M}_\text{t}$, we use the maximum $D$ reached in the outer wind.

\begin{figure}
	\centering
    \includegraphics[width=\hsize]{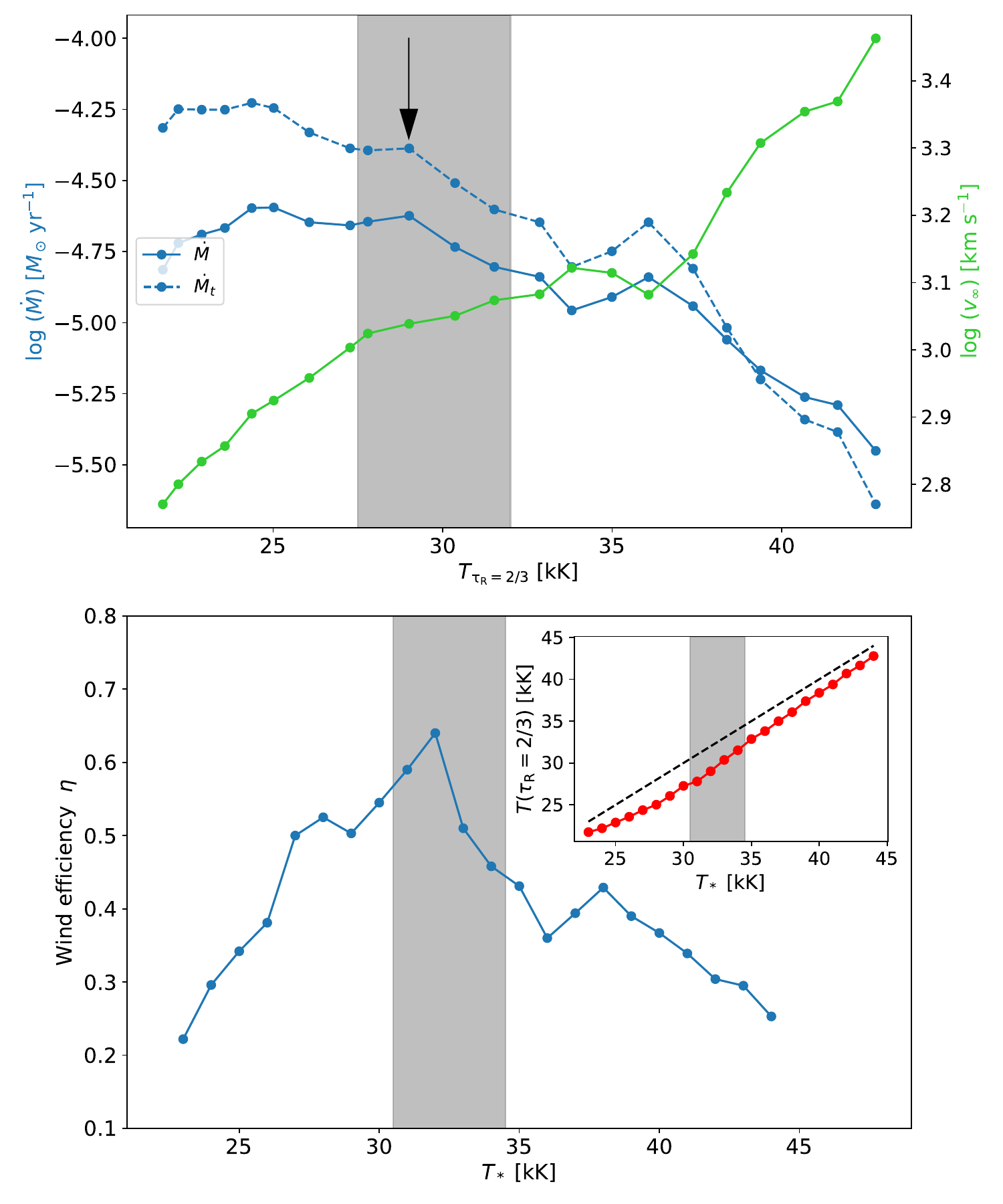}
	\caption{Mass-loss rates $\dot{M}$ and $\dot{M}_\mathrm{t}$ (top panel, blue dashed and blue solid lines, respectively), terminal velocities $\varv_\infty$ (top panel, in green) and wind efficiency parameters $\eta$ for the $T_\ast$ sequence of $Z_\ast = Z_\odot$ atmosphere models. 
    The shaded region denotes the transition regions of two  wind regimes.
    The arrow denotes the $T_\ast$ where $\eta$ overturns.}
	\label{fig:mdots_zsol}
\end{figure}

In Fig.\,\ref{fig:mdots_zsol}, we show the temperature trends for all four quantities. 
In the upper panel, $\dot{M}$, $\varv_\infty$, and $\dot{M}_\text{t}$ are plotted as functions of $T_{2/3}$, which is the resulting effective temperature at a Rosseland optical depth of $2/3$. 
Unlike $T_\ast$, which is an input parameter of the models, $T_{2/3}$ is an output parameter of the models. 
For models with strong winds, $T_\ast$ and $T_{2/3}$ can differ significantly \citep[see, e.g., Fig.\,6 in][]{sander2023}, but as we show in the inlet of the lower panel in Fig.\,\ref{fig:mdots_zsol}, the comparably moderate mass-loss rates of our models lead to an almost linear correlation of the two effective temperatures with $T_{2/3}$ being between $1$ and $3\,$kK lower than $T_\ast$\footnote{Our choice of $\tau_\text{cont,max} = 5$ compared to the more common values of $20$ and $100$ also reduces the difference a bit.}.

From the study of WN wind models launched by the hot iron bump \citep[cf.][their Figs.\,5, 6, and 10]{sander2023}, one might expect a smooth downward trend for $\dot{M}$ and $\dot{M}_\mathrm{t}$. 
This is not the case for our model sequence. 
While $\dot{M}$ shows the expected general downward trend with increasing temperature for $T_{2/3} \gtrsim 29\,$kK, there is a flattening for lower temperatures and eventual turnover of the trend for $T_{2/3} < 24\,$kK. 
In addition, there is a lot more substructure in the curve than seen in the model sequences from \citet{sander2023}.
The $\varv_\infty$-values across the temperature sequence are increasing overall, with a seemingly steeper increase of $T_\ast \geq 33$ kK. 
There is a notable exception for $34 \leq T_\ast \leq 36$ kK, with local decreasing $\varv_\infty$-values. 
Nonetheless, in this regime $\dot{M}$ increases when $\varv_\infty$ decreases and vice versa.
This is not the case for the regime below $25\,$kK, where $\varv_\infty$ keeps decreasing despite $\dot{M}$ decreasing as well.

When considering the transformed mass-loss rate $\dot{M}_\mathrm{t}$, the parallel decrease of $\dot{M}$ and $\varv_\infty$ below $25\,$kK effectively lead to a constant trend in $\dot{M}_\mathrm{t}$, which otherwise decreases with hotter temperature. 
The combined effect of $\dot{M}$ and $\varv_\infty$ further lead to an overturn for $T_\ast \lesssim 32$ kK in the wind efficiency $\eta$, which never manages to pass unity in our model sequence.
The trend changes are highlighted as grey areas in Fig.\,\ref{fig:mdots_zsol}.

\subsection{Line Driving}\label{subsec:line-driving}

\begin{figure}
	\centering
    \includegraphics[width=\hsize]{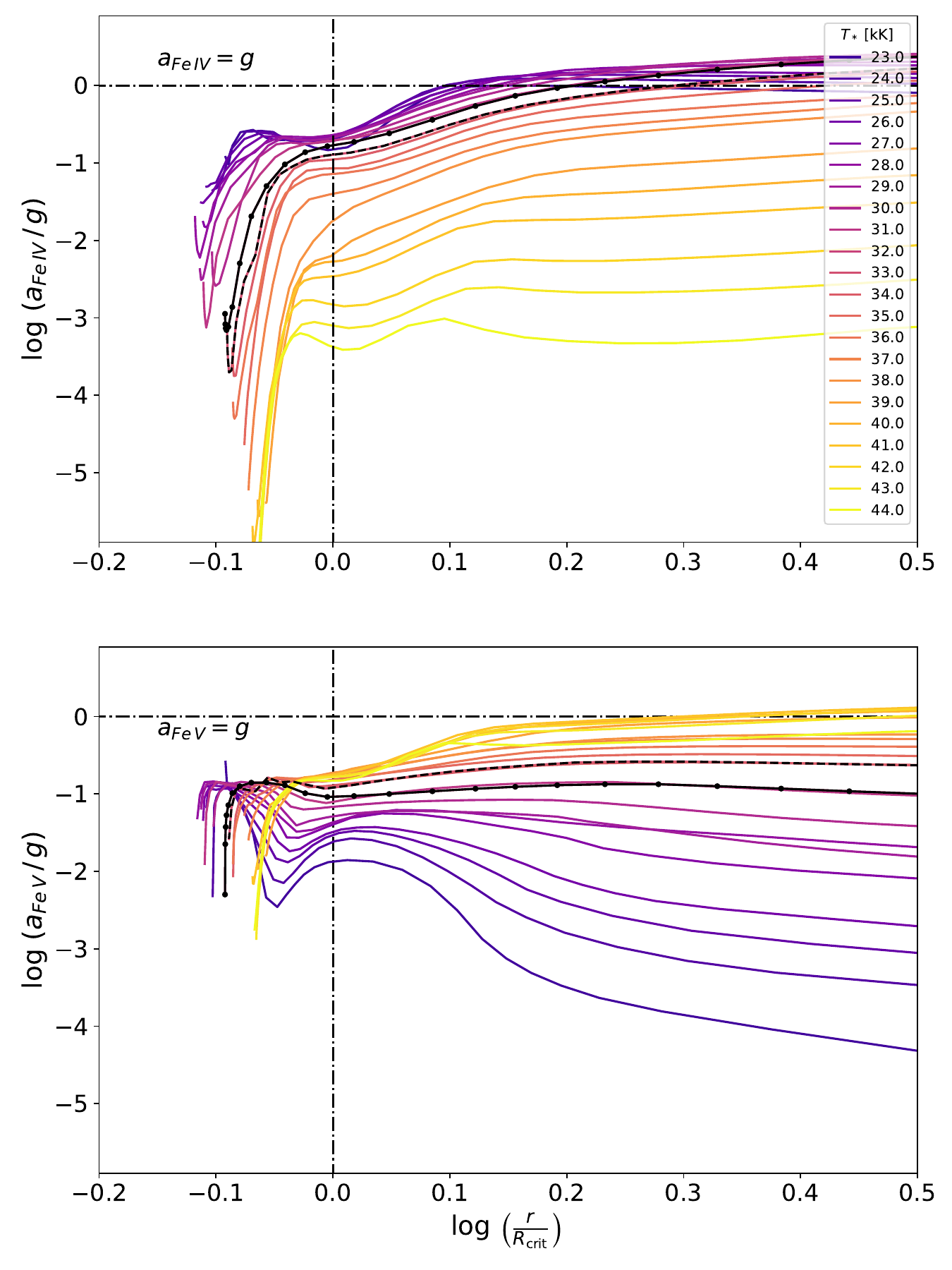}  
	\caption{
    Individual radiative accelerations throughout the wind of \ion{Fe}{iv} (top panel) and \ion{Fe}{v} (bottom panel), $a_\mathrm{\ion{fe}{iv}}$ and $a_\mathrm{\ion{fe}{v}}$ for the $Z_\ast = Z_\odot$ sequence. 
    Both accelerations are normalised by $g$. 
    The models at the turnover in Fig. \ref{fig:mdots_zsol} are highlighted; the $T_\ast=32$ kK model is solid dotted, $T_\ast = 33$ kK is dashed.}
	\label{fig:rad-acc_feiv-v_zsol}
\end{figure}

To gain a better understanding of the underlying mechanism of the global wind parameter trends and their changes, we examine the force contributions to the total wind acceleration in our model sequence. 
In general, our $L/M$-ratio is quite high with the resulting Eddington parameter\footnote{$\Gamma_\text{e}$ is defined similarly to the aforementioned $\Gamma_\text{rad}$, but takes only the acceleration due to (Thomson) electron scattering opacity into account.} $\Gamma_\text{e}(R_\ast) \approx 0.53$ for all of our models, meaning that at least at the inner boundary of the models the free electron opacity already accounts for half of the necessary force to overcome gravity. 
Still, line opacities are needed to fully launch the stellar wind. 
Similar to the models in \citet{sander2020driving,sander-vink2020}, or \citet{sander2023}, iron is the leading element for the line acceleration, which seems to be common for WR winds, at least at sufficiently high metallicity (cf.\, Sect.\,\ref{subsec:metallicity} and \ref{subsec:discussion-lowZ}).
However, the wind in our models is not launched by the iron M-shell ions of the hot iron bump (\ion{Fe}{vi}-\ion{Fe}{xvi}), but in a region where lower ionization stages are dominating, namely either \ion{Fe}{iv} or \ion{Fe}{v}. 
In Fig. \ref{fig:rad-acc_feiv-v_zsol}, we show the acceleration contributions (normalized by $g$) of $a_\mathrm{\ion{Fe}{iv}}(r)$ and $a_\mathrm{\ion{Fe}{v}}(r)$ for all models of the $Z_\odot$-sequence, normalised to $R_\text{crit}$ for an easier comparison. 
On first sight, the line accelerations show the expected behavior: the contribution of $a_\mathrm{\ion{Fe}{iv}}$ decreases for the hotter models while $a_\mathrm{\ion{Fe}{v}}$ increases due to the additional ionization depopulating \ion{Fe}{iv} in favor of \ion{Fe}{v}. 
A closer inspection reveals the model sets split up into two groups, which will become even more apparent when considering additional metallicities below. 
For $Z=Z_\odot$, the cooler models with $T_\ast \leq 32$ kK show little $a_\mathrm{\ion{Fe}{iv}}$ variation at $R_\mathrm{crit}$ and beyond (even when considering Fig.\,\ref{fig:rad-acc_feiv-v_zsol} showing a logarithmic scale), with a steep decrease for the hot model side, $T_\ast \geq 33$ kK. 
The reverse effect is true for $a_\mathrm{\ion{Fe}{v}}$.

For the model where the trend in $\eta$ changes (cf.\ Fig.\,\ref{fig:mdots_zsol}), there is a ``gap'' in the line acceleration stratifications shown in Fig. \ref{fig:rad-acc_feiv-v_zsol}.
In contrast, the region around  $34 \leq T_{2/3} \leq 36$ kK where the $\dot{M}$ trend deviates from the overall decline does not leave a clear imprint in the $a_\mathrm{\ion{Fe}{iv}}$ stratifications. 
There is also a small gap in the outer region of the or $a_\mathrm{\ion{Fe}{v}}$ stratification. 
Moreover, the bump in the $\dot{M}$-trend could be tied to a kink-like feature in the $a_\mathrm{\ion{Fe}{v}}$ stratification which gets close to $R_\text{crit}$ in this regime. 
For the hottest model, the $a_\mathrm{\ion{Fe}{v}}$-impact also goes down notably, explaining the steep decrease between the hottest and second-hottest model in Fig. \ref{fig:rad-acc_feiv-v_zsol}. 
Yet, despite all the features, the changes in the iron acceleration for the $Z_\odot$-models are rather smooth.

\subsection{Spectral imprint}\label{subsec:spectral}

\begin{figure}[h]
	\centering
    \includegraphics[width=\hsize]{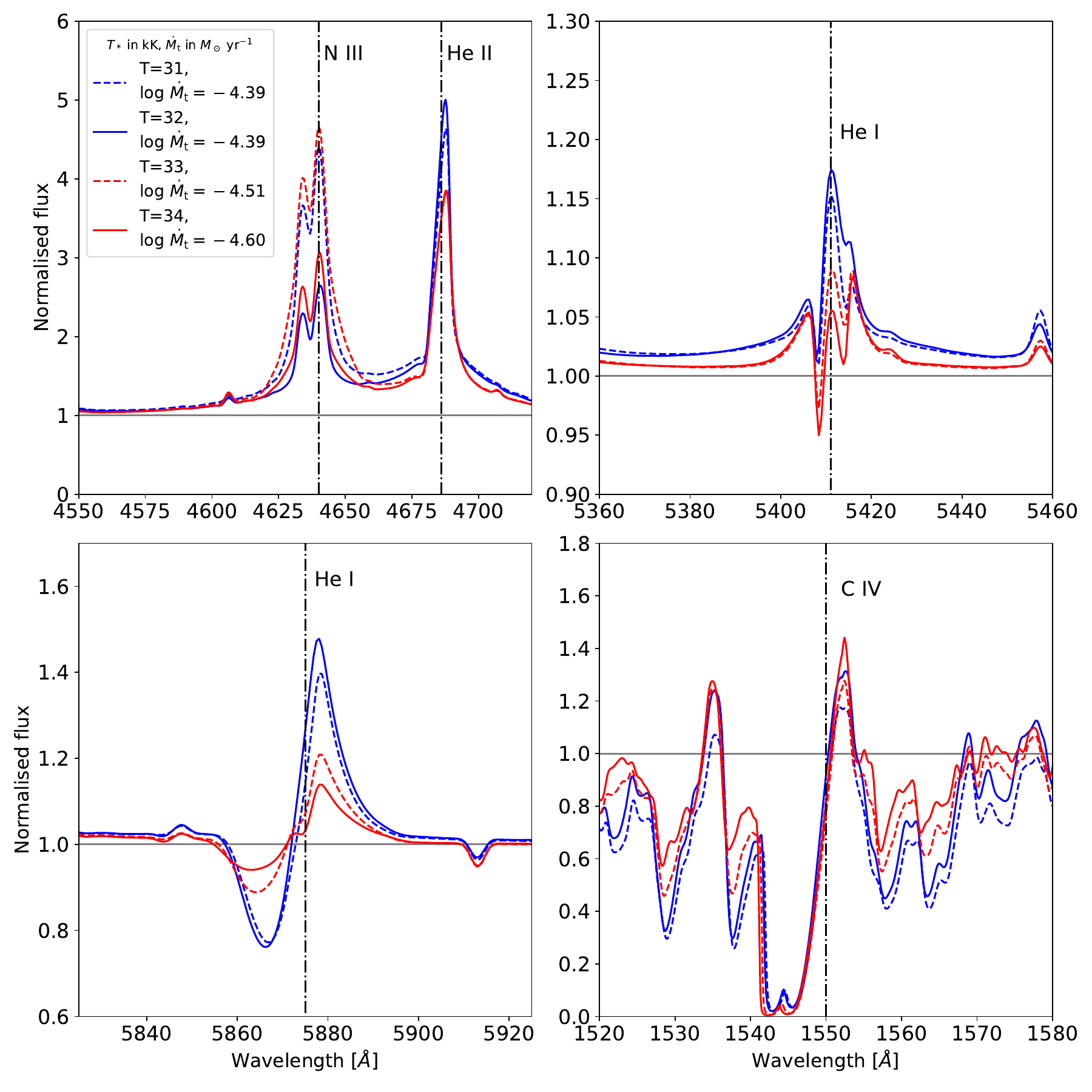} 
	\caption{A selection of optical diagnostic lines commonly used for WN-star spectral type determination \citep[see][]{smith1996} (top panels and bottom left panel), and the \ion{C}{iv} $\lambda\lambda\,1550$ \AA P Cygni profile.
    The four selected model spectra go over the turnover shown in Fig. \ref{fig:mdots_zsol}.
    }
	\label{fig:diagnostics-zsol}
\end{figure}

The spectral appearance of WR stars is significantly affected by the wind stratification. 
Even in the case of $\beta$-type velocity laws, the effect on the choice of $\beta$ for the observed emission lines can be significant \citep[see, e.g.,][]{lefever2023}. 
In Fig.\,\ref{fig:diagnostics-zsol}, we show example spectra from our $Z_\odot$-sequence for models with $T_\ast = 31, 32, 33\, \mathrm{and}\, 34$ kK, i.e., over the region where $\eta$ turns over (see also Fig. \ref{fig:mdots_zsol}). 
Including some of the diagnostic lines from the WN-star classification by \citet{smith1996}, we plot the optical \ion{N}{iii} $\lambda\,4640$ \AA, \ion{He}{ii} $\lambda\,4686$ \AA, \ion{He}{i} $\lambda\,5411$ \AA, and \ion{He}{i} $\lambda\,5875$ \AA, as well as the \ion{C}{iv} $\lambda\lambda\,1550$ \AA\ UV line to illustrate the effect on P\,Cygni lines.
From the raw temperature effect on the spectra, one expects the spectral lines to show some gradual changes over this 3 kK range. 
Due to the inherent coupling with different wind parameters, the spectra actually display a sharper difference between $T_\ast = 32$ and 33\,kK. 
This is especially clear for the \ion{He}{i} and \ion{He}{ii} lines in Fig.\,\ref{fig:diagnostics-zsol}. 
Still, the hotter models show weaker \ion{He}{i} lines as expected.
In contrast, the \ion{N}{iii} $\lambda\,4640$ \AA\, lines show a different behavior with the 31 and 33\,kK models being similar to each other on the one hand and the 32 and 34\,kK models on the other hand. 
As discussed in detail in \citet{Rivero-Gonzalez+2011}, these 3d$\rightarrow$3p nitrogen-lines have a complex origin. 
In O-stars, these lines can originate in the (quasi-)hydrostatic layers and still be in emission due to non-LTE effects. 
Depending on the model assumptions and the atomic data treatment, the line predictions can change considerably \citep[see, e.g., the recent comparison in][]{Sander+2024}. 
For WR stars, they eventually become wind lines, but are still very sensitive to the velocity and thus density. 
As reflected by our $\eta$ and $\dot{M}_\text{t}$ values, the winds of our models are not that dense, despite the nominally high mass-loss rates. 
Hence, even if the $\dot{M}$ and $\varv_\infty$ changes between two models are moderate, the exact hydrodynamic solution in the wind onset region can overturn the gradual temperature effects on the spectral lines. 
This is different for the UV \ion{C}{iv} $\lambda\lambda\,1550$ \AA\, line, where the absorption is produced in the outer wind and we see only gradual changes, confirming that the transitions of the global wind behavior are rather smooth for $Z_\odot$.

\subsection{Metallicity influence}\label{subsec:metallicity}

\begin{figure}[h]
	\centering
    \includegraphics[width=\hsize]{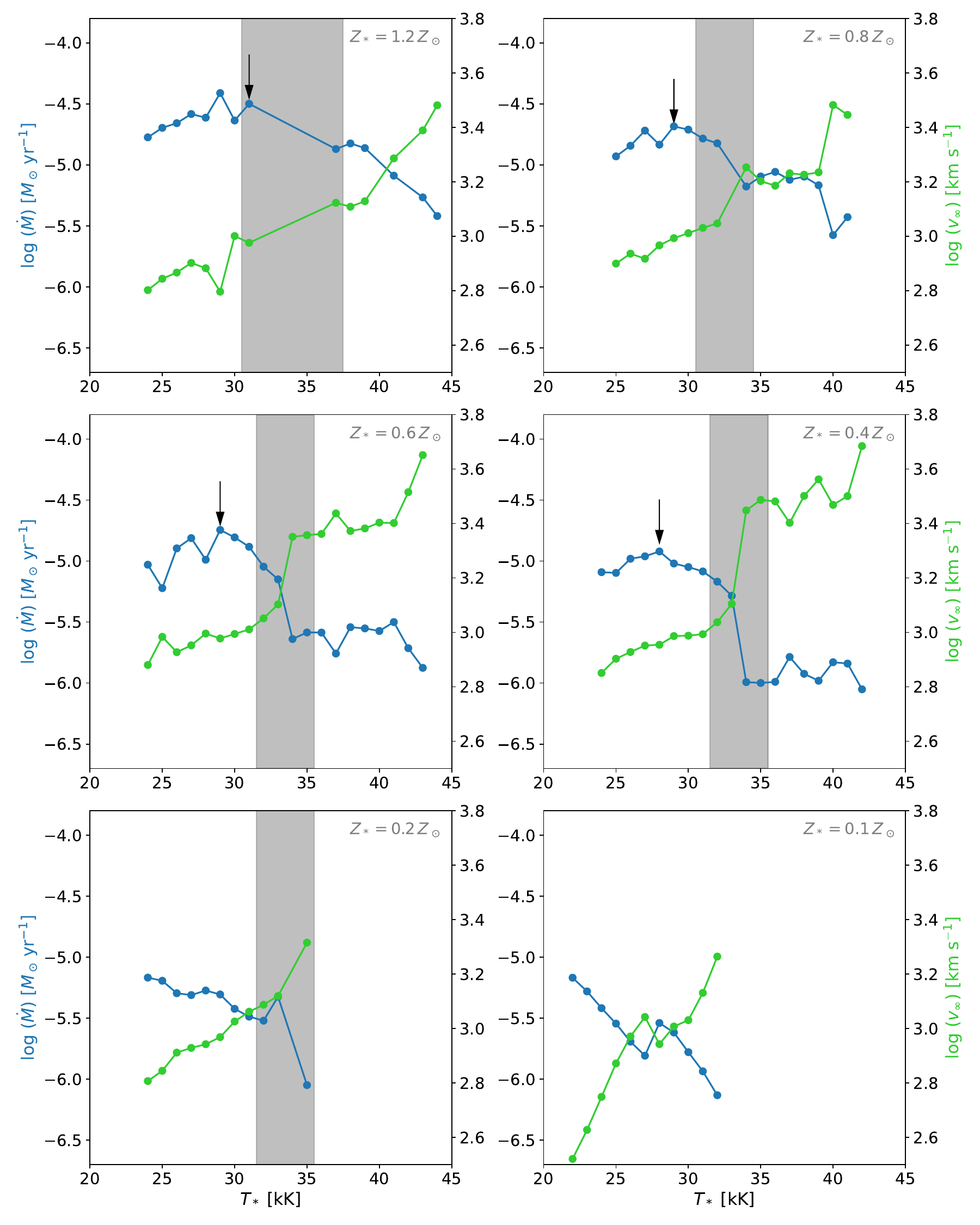} 
	\caption{Similarly to Fig. \ref{fig:mdots_zsol}, but now for other $Z_\ast$ sequences (denoted in grey on the top right of each panel). 
    The $\dot{M}$ values are in blue, the $\varv_\infty$ values are in green; the shaded regions and arrows denote the wind regime change and the turnover in $\eta$ respectively, if it is present.}
   \label{fig:mdots-vfinals-zs}
\end{figure}

After examining the $Z_\odot$-sequence, we now study additional model sequences at different metallicities, ranging from 1.2 to 0.02 $Z_\odot$ (see Table\,\ref{tab:model-params} for an overview of the models per temperature and metallicity). 
The $\dot{M}$ and $\varv_\infty$-values of all temperature sequences except solar are shown in Fig.\,\ref{fig:mdots-vfinals-zs}, similarly to the top panel in Fig. \ref{fig:mdots_zsol}. 
Iron driving and spectral imprint plots similar to Figs.\,\ref{fig:rad-acc_feiv-v_zsol} and \ref{fig:diagnostics-zsol} can be found in appendices \href{https://doi.org/10.5281/zenodo.15675082}{B} and \href{https://doi.org/10.5281/zenodo.15675082}{C}, respectively. 

\paragraph{$1.2\,Z_\odot$:}
At super-solar metallicity, we see a generally similar behavior to the solar metallicity case (cf.\,\ref{fig:mdots-vfinals-zs}): there is an overturn in $\dot{M}$, albeit now in ``double-peak'' form and a flattening of $\dot{M}_\mathrm{t}$ for cool temperatures. 
On the hot side, again a strong decrease with temperature is obtained.
As in the $Z_\odot$ case, we can distinguish two different wind regimes based on the difference in the $a_\mathrm{\ion{Fe}{iv}}$ and $a_\mathrm{\ion{Fe}{iv}}$ trends, shown in the left-hand panel of Fig.\,\href{https://doi.org/10.5281/zenodo.15675082}{B1}. 
The exact point of the transition between the regimes is less clear as these models are close to the Eddington limit and the radiative force in this (even moderately) supersolar regime is so high, that models with $T_\ast$ between 32 to 36 kK did not converge within the same model setup.
Based on the results of the $Z_\odot$ models and the wind efficiency of the $1.2\,Z_\odot$ models peaking at 31 kK, we expect the transition to happen similarly smoothly, now at $T_\ast \approx$ 31-32 kK.

\paragraph{$0.8\,Z_\odot$:}
A different kind of behavior occurs when transitioning to sub-solar metallicity. 
A clear trend discontinuity (``jump'') occurs between $T_\ast = 32$ and $33$\,kK, both in $\dot{M}$ and $\varv_\infty$.
The changes in $a_\mathrm{\ion{Fe}{iv}}$ and $a_\mathrm{\ion{Fe}{v}}$ (cf.\ top-right panel in Fig. \href{https://doi.org/10.5281/zenodo.15675082}{B1}) are now very distinct over this jump, contrary to $Z_\odot$.
Moreover, at $Z_\odot$, the trend in $\dot{M}_\text{t}$ started to flatten where $\eta$ peaks, which also  aligned with the iron switch. 
This is not the case in the $0.8\,Z_\odot$ sequence, where $\eta$ peaks at $29\,$kK, i.e., below the iron switch. 
Similar to $Z_\odot$, however, $\dot{M}$ peaks where $\eta$ peaks and $\dot{M}_\text{t}$ approximately flattens for cooler temperatures.\\
The sharp drop between $T_\ast = 32$ and $33$\,kK also results in $\eta$ decreasing by a factor of two. 
Interestingly, at $T_\ast \geq34$ kK, the wind regime seems to ``recover'' to the trend started on the cool side of the jump, to then drop again at for models where $T_\ast\geq 40$ kK.

\paragraph{$0.6\,Z_\odot$:}
When decreasing the metallicity further, we see a sharp jump in the $\dot{M}$- and $\varv_\infty$ trends (see the left-middle panel in Fig.\,\ref{fig:mdots-vfinals-zs}), now occurring between $T_\ast = 33$ and 34\,kK. 
Below the jump, $\dot{M}$ is approximately constant before eventually decreasing again.

\begin{figure}[h]
	\centering
    \includegraphics[width=\hsize]{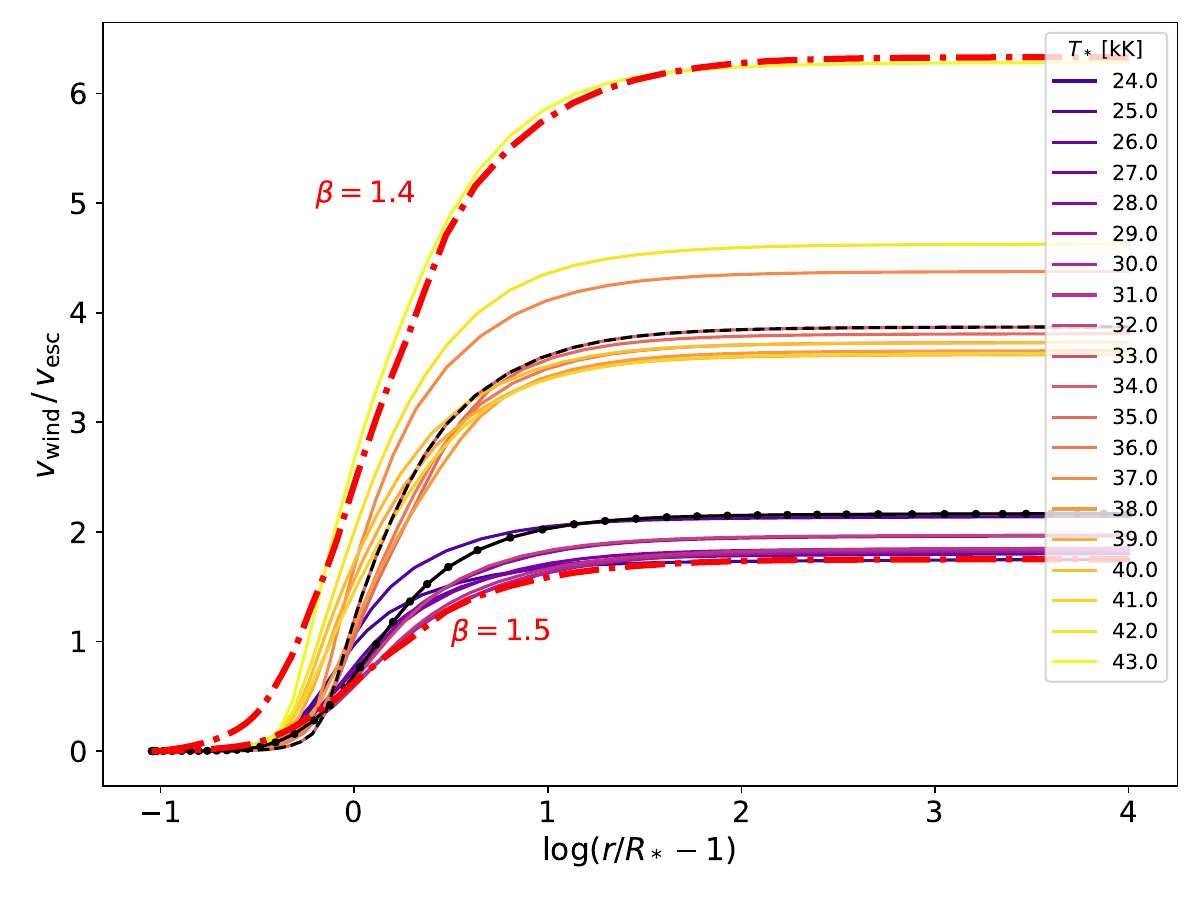} 
	\caption{Velocity fields for the $0.6\,Z_\odot$ temperature sequence. 
    Similarly to Fig. \ref{fig:rad-acc_feiv-v_zsol}, the models adjacent to the sharp wind transition are shown; dotted solid for the $T_\ast=33$ kK model, dashed for the $34$ kK model.
    Two $\beta$ velocity laws are shown with $\beta$ such that they fit closest to the coolest and hottest models.}
	\label{fig:vfield-0.6zsol}
\end{figure}

This behavior shows similarities to the bi-stability jump in the B supergiant models from \citet{vink1999}. 
The wind velocity stratification now clearly shows two separates regimes as depicted in Fig.\,\ref{fig:vfield-0.6zsol}. 
This did not occur for $1.2\,Z_\odot$ and $Z_\odot$ metallicities. 
For the $0.8\,Z_\odot$ sequence, the situation is less clear.
The jump in the $0.6\,Z_\odot$-sequence coincides with differences in the $a_\mathrm{\ion{Fe}{iv}}$ and $a_\mathrm{\ion{Fe}{v}}$ stratifications (cf.\ the lef panel of Fig.\,\href{https://doi.org/10.5281/zenodo.15675082}{B2}). 
Yet, the position of the jump does not exactly coincide with \ion{Fe}{iv} taking over from \ion{Fe}{v} at $R_\text{crit}$. 
On both sides, \ion{Fe}{v} is the leading line contribution at $R_\text{crit}$ and although the importance of \ion{Fe}{iv} increases more and more, it only becomes the lead line acceleration at $30\,$kK and below. 
At $29\,$kK, $\dot{M}$, $\eta$ and $\dot{M}_\text{t}$ reach their peak values in the $0.6\,Z_\odot$-sequence. 
The regime below shows a bit of scatter in the $\dot{M}$ and $\dot{M}_\text{t}$-trend. 

\begin{figure}
    \centering
    \includegraphics[width=\hsize]{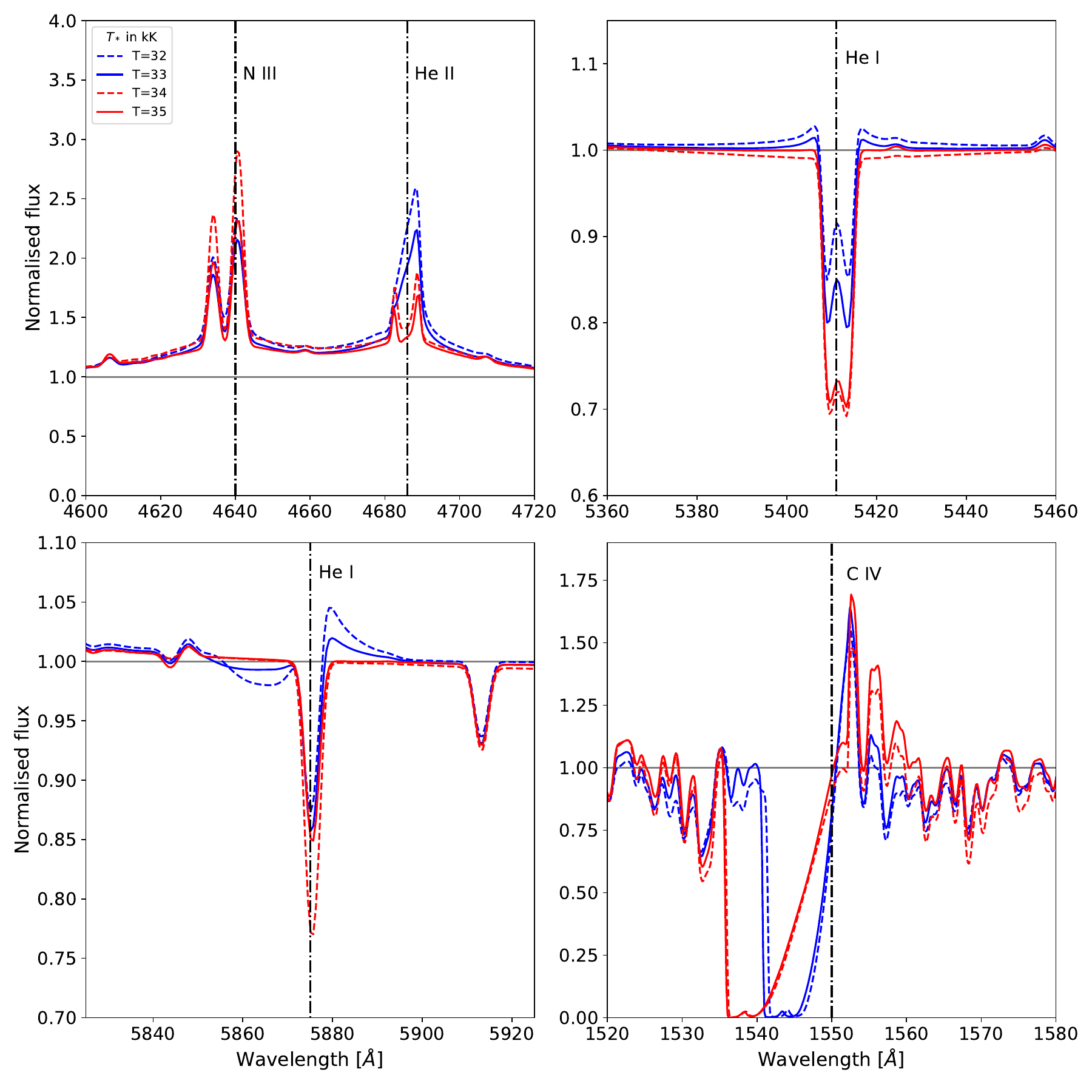}
    \caption{Similar to Fig. \ref{fig:diagnostics-zsol}, but here for the $Z = 0.6\,Z_\odot$ sequence and with models over the sharp wind transition.
    An overview of the same models in the K-band is shown in Fig. \href{https://doi.org/10.5281/zenodo.15675082}{C1}}
    \label{fig:diagnostics-0.6zsol}
\end{figure}

In Fig.\,\ref{fig:diagnostics-0.6zsol}, we show the corresponding spectra across the jump between 33 and 34\,kK. 
The changes in $\dot{M}$ are now clearly visible in the lines (except for the aforementioned \ion{N}{iii} lines) while the change in $\varv$ is only evident in the UV.

\paragraph{$0.4\,Z_\odot$:}
The behavior of the $0.4\,Z_\odot$-sequence is very similar to the $0.6\,Z_\odot$ sequence. 
Again, there is a sharp jump in $\dot{M}$ and $\varv_\infty$ between 33 and 34\,kK and an approximately constant $\dot{M}$ for the temperatures below as well as clear stratification differences in $\varv(r)$.
The similarities can also be noticed in the $a_\mathrm{\ion{Fe}{iv}}$ and $a_\mathrm{\ion{Fe}{v}}$ stratification (see left panel of Fig.\,\href{https://doi.org/10.5281/zenodo.15675082}{D2} and right panels of Fig.\,\href{https://doi.org/10.5281/zenodo.15675082}{B2}) and the spectral imprint (see Fig. \href{https://doi.org/10.5281/zenodo.15675082}{C5}).
The $\eta$ turnover now occurs at $28\,$kK, but is otherwise in a similar fashion as seen for the 0.8 and 0.6 $Z_\odot$ sequences.

\paragraph{$0.2$ to $0.02\,Z_\odot$:}
At the lowest metallicity values studied in this work, the hotter models hardly converge to a solution within the given framework. 
For example, we find an upper limit of $-6.48$ for the $36\,$kK model at $0.2\,Z_\odot$, but the model would not be considered converged by the same criteria as the other models. 
Showing only those models that converged within the same set of criteria in Fig.\,\ref{fig:mdots-vfinals-zs}, we can only resolve the beginning of the jump at $0.2\,Z_\odot$ (between 33 and 35\,kK). 
The jump heavily influences the wind stratification (see Figs.\,\href{https://doi.org/10.5281/zenodo.15675082}{D2} and \href{https://doi.org/10.5281/zenodo.15675082}{B3}) and the spectra (see Fig.\,\href{https://doi.org/10.5281/zenodo.15675082}{C5}), similar to the $0.6\,Z_\odot$ and $0.4\,Z_\odot$ sequences. 
On the hotter side, the trend in $\eta$ no longer turns over, but flattens for $T_\ast \leq 29\,$kK. This is a consequence of $\dot{M}$ no longer decreasing for the coolest models. 

For the $0.1\,Z_\odot$ to $0.02\,Z_\odot$ sequences, the general picture is similar to the $0.2\,Z_\odot$-sequence with iron becoming even less important. Within our covered $T_\ast$ regime, we do no longer find a jump. 
While there is a some discontinuity in the $\dot{M}$ trend between $27$ and $28\,$kK for $0.1\,Z_\odot$, there is no dramatic drop of $\dot{M}$, but a rather continuous downward trend in the temperature range where we found the jump at higher $Z$.
There is no local maximum in $\dot{M}$, but the $\eta$-values seem to settle for the cool temperature end of our sequence.
As we will discuss further in Sect.\,\ref{sec:discussion}, the main driving ingredients in this low-metallicity regime change. 
The wind-launching opacity bump formed by \ion{N}{iii} and \ion{N}{iv} is further supported by other elemental ions, most notably \ion{S}{v} and \ion{O}{iv}. 
Carbon hardly plays a role due to the assumed WN-type composition where most carbon has been transformed into nitrogen.

%-----------------------------------------------------------------
\section{Discussion}\label{sec:discussion}

\subsection{The turnover of $\dot{M}$ at low temperatures and higher metallicities}
\label{subsec:discussion-turnover}

In all of the sequences between $1.2\,Z_\odot$ and $0.4\,Z_\odot$, we find that $\dot{M}$ and thus also $\eta$ do not continue to monotonously increase with lower $T_\ast$, but reach a maximum and eventually decline again. 
A deeper inspection of the radiative accelerations of all models reveals that this is most likely related to a small opacity peak resulting from the \ion{Fe}{v} to \ion{Fe}{vi} recombination. 
At the hot side of the turnover, this opacity bump is located right below the wind launching $R_\text{crit}$ and thus helps to boost the mass-loss rate. 
For cooler temperatures, this bump moves inwards, now slightly beneath the wind launching region. 
There, the peak is not sufficient to launch the wind in our current framework, but affects the solution depending on whether or not a small super-Eddington regime has to be suppressed. 
Above this bump, a ``dip'' occurs in the \ion{Fe}{iv} acceleration which leads to a reduction in $\dot{M}$. 
A similar effect is observed for OB-star winds in \cite{muijres2011} and \cite{vink&sander2021}
The magnitude of the dip is also connected to the scatter in $\dot{M}$ in the $T_\ast$ regime below the maximum. 
Eventually, $\dot{M}$ is expected to rise again when the re-increase of the \ion{Fe}{iv} acceleration reaches $R_\text{crit}$, but this does not yet happen for our coolest models in the sequences.

Whether this overturn in the mass-loss rate is realized in nature, is an open question. As demonstrated by \citet{moens2022}  and \citet{debnath2024}, radiatively-driven turbulence could occur for a near- or super-Eddington opacity bump slightly below the stellar surface. As we demonstrate in \citet{Gonzalez-Tora+2025}, this can affect the wind-launching solution and thus might ``wash out'' the effects of the bump. Yet, studying this would require a depth-dependent implementation of turbulent pressure, which is beyond the scope of the current work.

For the lowest metallicities ($Z \leq 0.2\,Z_\odot$), the turnover does no longer occur. This is a natural consequence of the diminished importance of iron, which no longer dominates the wind solution.

\subsection{Strong discontinuities (``jumps'') in the $\dot{M}$-trend}\label{subsec:discussion-jumps}

\begin{figure}
    \centering
    \includegraphics[width=\hsize]{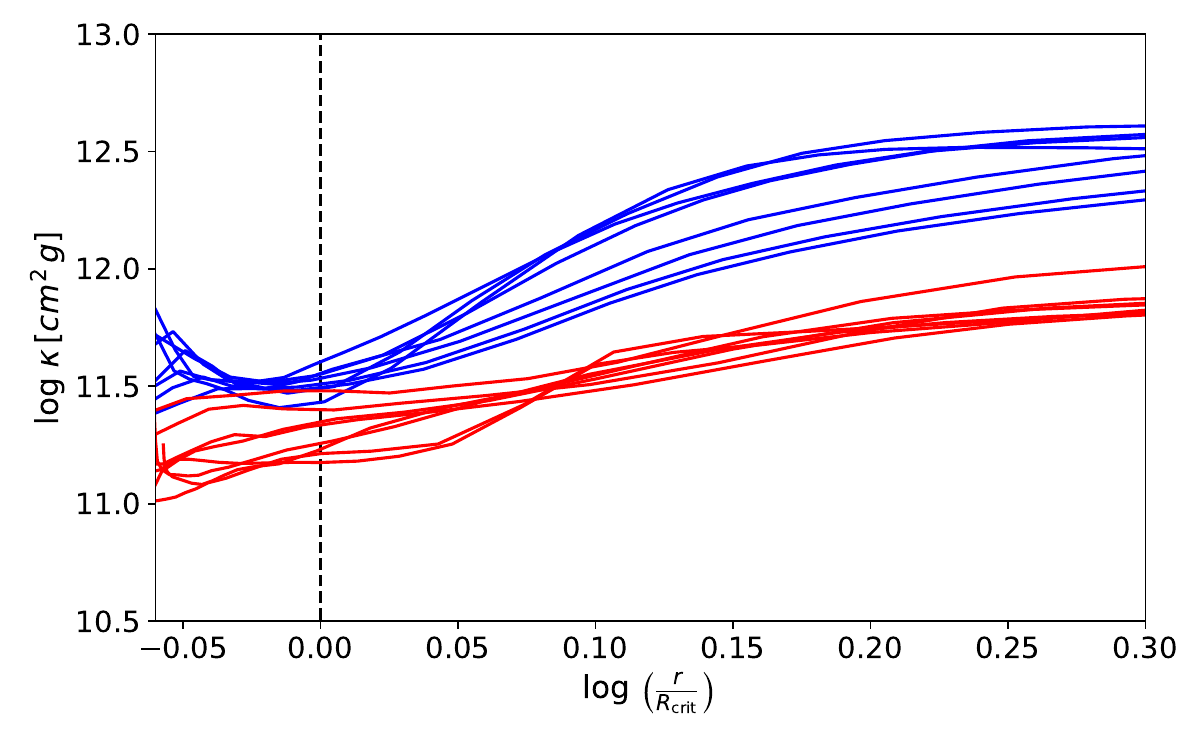}
    \caption{Flux-weighted mean opacities $\varkappa_F$ for iron ions of the $0.8\, Z_\odot$ sequence.
    The blue curves represent the blue side of the $\dot{M}$ jump in Fig. \ref{fig:mdots-vfinals-zs}, where the red curves represent the hot side.}
    \label{fig:opacities-0.8zsol}
\end{figure}

For several, but not all metallicities, we obtain a strong jump in the $\dot{M}(T_\ast)$ trend. Interestingly, the jump is not present at $Z \geq 1.0\,Z_\odot$, but between $0.8$ and $0.4\,Z_\odot$. 
At first, this might seem counter-intuitive due to the increasing role of line opacities at higher metallicities. 
While the jump occurs close to the \ion{Fe}{v} to \ion{Fe}{iv} recombination regime, the inspection of our models indicate that that is not an immediate consequence of this ionization switch. 
It becomes especially evident considering the $0.4\,Z_\odot$ sequence, where the leading acceleration does not change at the same $T_\ast$ where the jump in $\dot{M}$ occurs. 
Instead, the jump is the consequence of a change in the iron opacity behavior. 
On the cool side of the jump, where $\dot{M}$ is significantly higher, the flux ($F(r)$)-weighted mean iron opacity

\begin{equation}
   \varkappa_{F}(r) = \frac{1}{F(r)} \int\limits_{0}^{\infty} \varkappa_\nu(r) F_\nu(r)\,\mathrm{d}\nu
\end{equation}

increases outwards at $R_\text{crit}$, while this is not the case for the hotter models with the weaker mass-loss rate. 
As pointed out by \citet{Nugis2002}, $\varkappa_{F}$ must increase at $R_\text{crit}$ to get a stable wind solution. 
If the leading Fe ion opacity does not provide this, other, less efficient line opacities -- either from a different Fe stage or from an element like nitrogen -- will eventually do when the mass-loss rate is reduced. 
When both Fe ions do not help in contribution to a increasing opacity over $R_\text{crit}$ at a certain $T_\ast$, but they do for the next-cooler model, a large jump in $\dot{M}$ is obtained. 
This effect is showcased in Fig. \ref{fig:opacities-0.8zsol} in the case of the $0.8\,Z_\odot$ sequence.

In the high-metallicity cases where there is no jump (e.g., at $Z_\ast = Z_\odot$), this split-up in the iron opacity gradient into an increasing and non-increasing regime over $R_\text{crit}$ does not occur. 
However, the $\varkappa_{F}$-values keeps increasing outwards for all models. 
For the lowest metallicities on the other hand ($Z \leq 0.2\,Z_\odot$), the role of iron is no longer important enough to dominate the wind solution. 
Nitrogen instead becomes the leading line opacity and shows an outward-increasing $\varkappa_{F}$ around $R_\text{crit}$ for the whole model sequence, thereby preventing any strong discontinuities in $\dot{M}$.
In \cite{vink2001}, carbon takes over from iron as the main driver at low metallicity; we do not observe minor jump behaviour due to (e.g., carbon) recombination effects in the low metallicity case however.

Currently, it is unclear whether the obtained sharp discontinuities could potentially be smeared out due to multi-dimensional effects. 
In current multi-D simulation efforts \citep[e.g.,][]{moens2022,debnath2024}, significant fluctuations of the temperature are seen throughout the wind, reaching differences of $10\,$kK and more. 
Assuming this would -- very roughly -- correspond to an averaging between different 1D $T_\ast$ solutions, we cannot rule out that our obtained jumps in the $\dot{M}$ trend could be considerably reduced by multi-D effects.

\subsection{WNh wind driving at very low metallicities}\label{subsec:discussion-lowZ}

\begin{figure}
    \centering
    \includegraphics[width=\hsize]{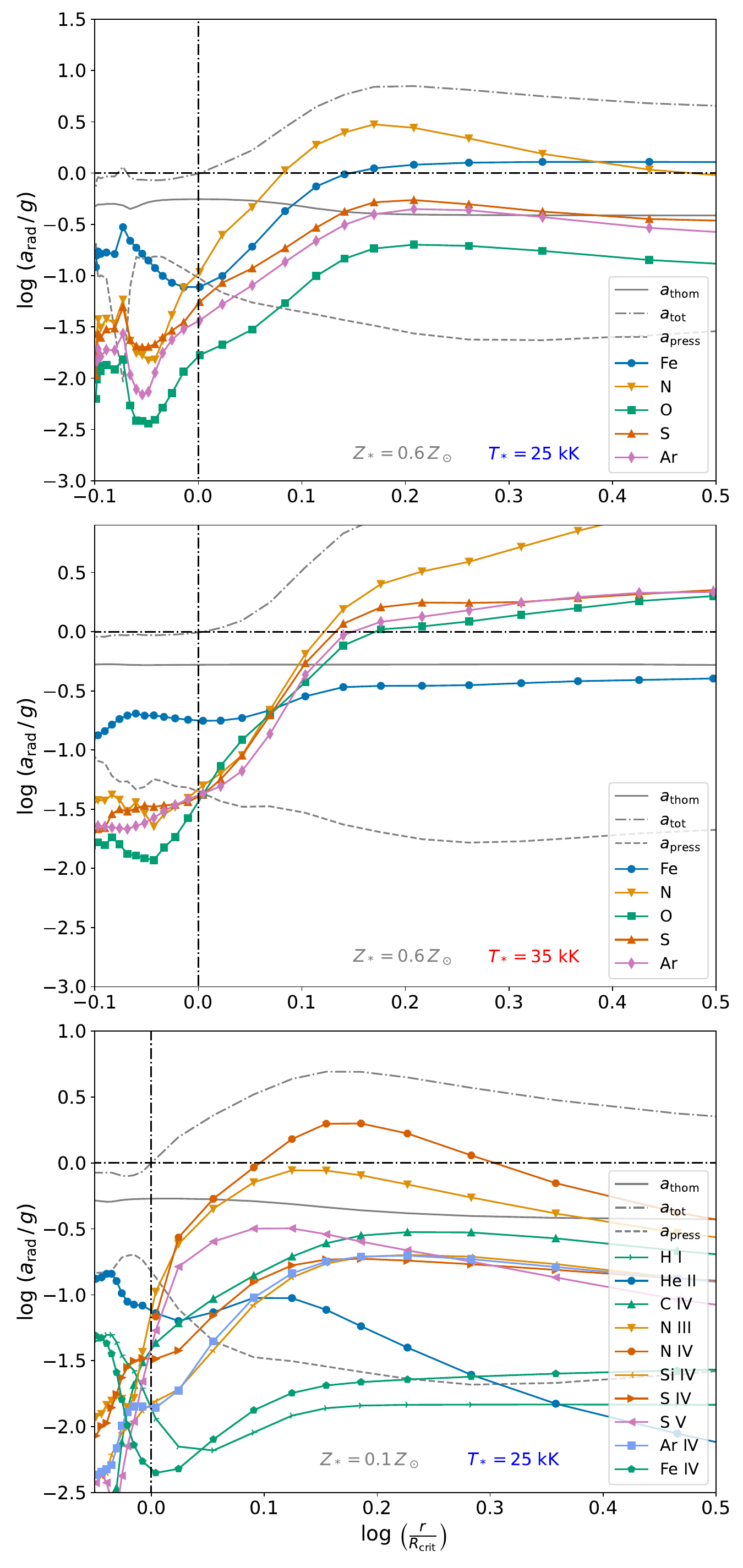}
    \caption{Top two panels: Major contributions to the total wind acceleration $a_\text{tot}$ for a hot (top) and cold (mid) model from the $0.6\,Z_\odot$ sequence. Major elemental contributions to the summed-up radiative acceleration $a_\mathrm{rad}$ are indicated as well.
    The bottom panel shows the $0.1\,Z_\odot$-model at $T_\ast = 25\,kK$, but now lists the strongest contributions from individual ions to $a_\text{rad}$.}
    \label{fig:radacc-comp}
\end{figure}

In all our models the recombination of \ion{He}{iii} to \ion{He}{ii} only happens beyond $R_\text{crit}$. 
Thus, despite the important role of iron, the Thomson free electron contribution actually outweighs the acceleration of any specific element at $R_\text{crit}$ at all metallicities. 
This is also illustrated in the three examples in Fig.\,\ref{fig:radacc-comp} showing the most important contributions to the total wind acceleration $a_\text{tot}$ in the wind onset region for three different models.
However, as $\Gamma_\text{e}(r)$ is rather smooth, the wind solutions are dominated by the radially more varying acceleration terms. 
At $Z \leq 0.2\,Z_\odot$ the role of the Fe opacities becomes so weak in our models, that eventually other opacity sources take over. 
This is illustrated in the bottom panel of Fig.\,\ref{fig:radacc-comp}, where we see that in particular nitrogen opacities of $\ion{N}{iii}$ and \ion{N}{iv} and -- at $T_\ast \lesssim 29\,$kK -- also the bound-free opacity from the continuum transitions of \ion{He}{ii} are the leading ion contributions at $R_\text{crit}$. \ion{S}{iv} and \ion{S}{v} ions also play a role as do individual ions of elements such as C, Si, or Ar.

\begin{figure}
    \centering
    \includegraphics[width=\hsize]{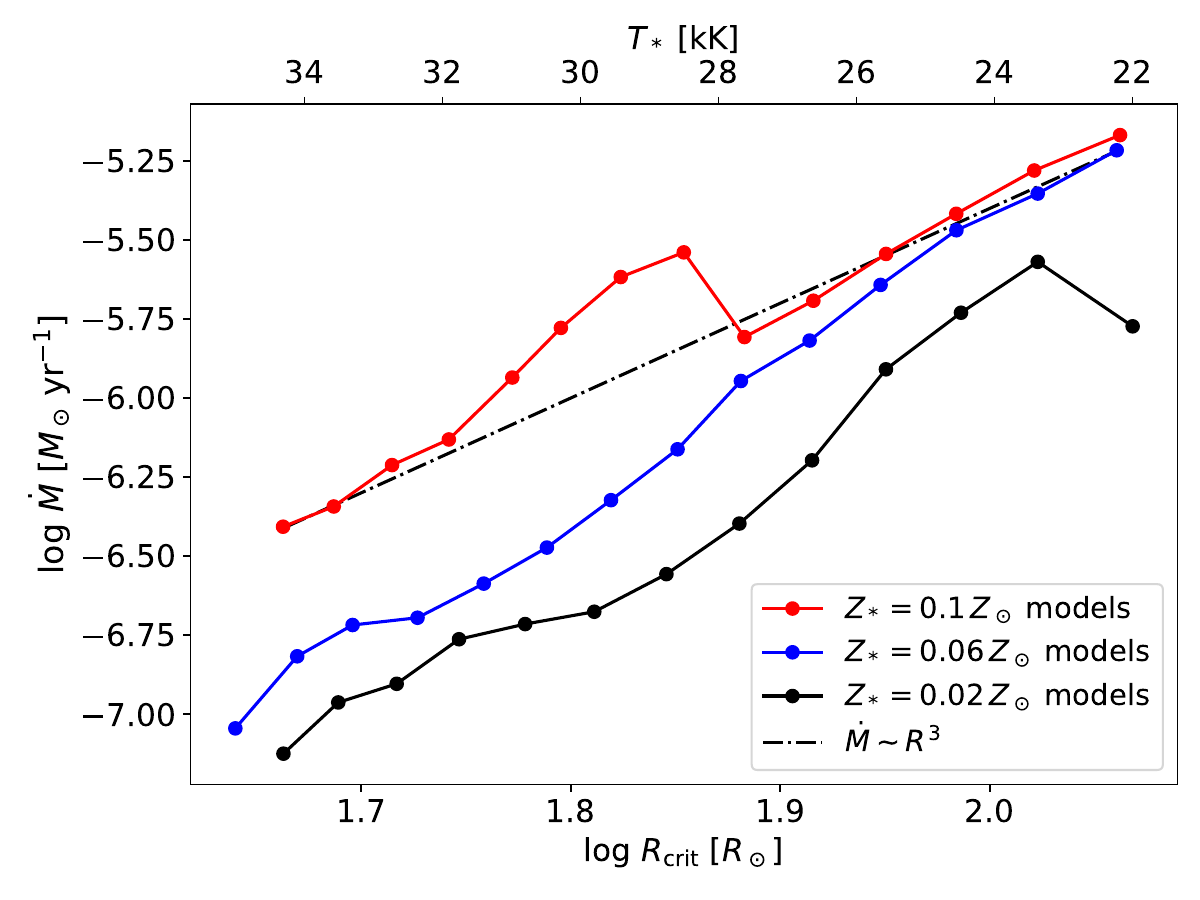}
    \caption{The $\dot{M}$-values are shown at the critical radii $R_\text{crit}$ for low-metallicity sequences of $0.1$, $0.06$, and $0.02\,Z_\odot$. 
    The dashed line represents the geometric effect of $\dot{M}\sim R^3_\mathrm{crit}$, fitted for the $0.1\,Z_\odot$ sequence.}
    \label{fig:r3-dependence}
\end{figure}

With the complex Fe opacity no longer setting the overall $\dot{M}$ trend, we find an increasing mass-loss rate with lower $T_\ast$, driven by the combined continuum and line accelerations mentioned above. 
Interestingly, this trend is much more smooth then we otherwise find in our sample. As illustrated in Fig.\,\ref{fig:r3-dependence}, in this regime we closely resemble the relation

\begin{equation}
   \dot{M} \propto R_\text{crit}^3
\end{equation}

found for the dense-wind regime of cWR stars in \citet{sander2023}. Albeit being a very different parameter regime, this means that the pure geometrical effect from the lower temperature, i.e., the implied lower $\log g(R_\text{crit})$ is the dominant reason for the $\dot{M}$ increase. 
Running a few models for even lower metallicities of $0.06\,Z_\odot$ and $0.02\,Z_\odot$ in this comparably higher mass-loss regime shows a similar trend, but shifted to lower $\dot{M}$ as expected by the further decreasing line opacities.

\begin{figure}
    \centering
    \includegraphics[width=\hsize]{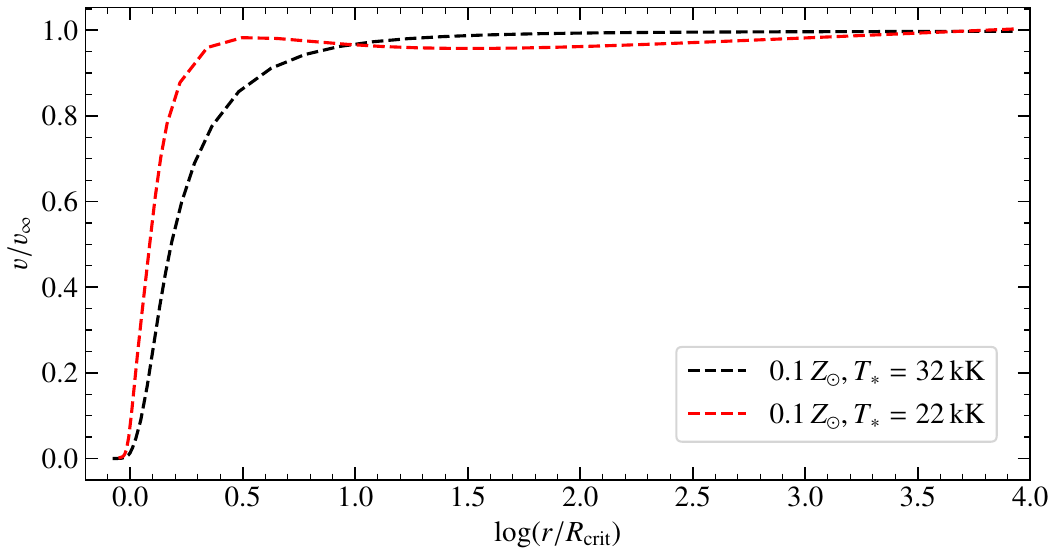}
    \caption{Different types of velocity fields obtained in the low metallicity regime. The hot model shows a typical radiation-driven wind settling on a terminal velocity, while the cool model shows a deceleration region and then a shallow, but ever-growing velocity.}
    \label{fig:velolowz}
\end{figure}

In the outer wind, different ions contribute to the wind, depending on $T_\ast$. 
For lower temperatures, \ion{Fe}{iv} still has an impact similar to carbon, nitrogen and sulfur there at $0.1\,Z_\odot$. For hotter models, iron hardly plays a role in the outer wind, but instead argon gets more important while nitrogen, carbon and sulfur keep their prominent role. 

For the coolest models and $Z \leq 0.1\,Z_\odot$, the terminal velocities become very low ($\varv_\infty \lesssim 500\,\mathrm{km\,s}^{-1}$). 
Here, the radiative acceleration alone would not be sufficient to sustain the outer wind, but here the slow-falling $1/r$-term from the gas pressure contribution helps to keep the material escaping. 
This is also evident from the resulting velocity fields shown in Fig.\,\ref{fig:velolowz}, where we obtain a non-monotonic solution from the hydrodynamic equation of motion for the cool model.\footnote{For the radiative transfer calculation, the non-monotonic part in $\varv(r)$ is interpolated as explained and illustrated in \citet{sander2023}.}. 
A similar dichotomy is obtained in models with even lower metallicity of $0.06\,Z_\odot$ and $0.02\,Z_\odot$ (in the same $T_\ast$ region). With decreasing $Z$, $\varv_\infty$ first increases due to the effect of the decreasing $\dot{M}$. 
Eventually though, the opacity in the wind to support higher $\varv_\infty$ also decreases and thus $\varv_\infty$ goes down again. 

\begin{figure}
    \centering
    \includegraphics[width=\hsize]{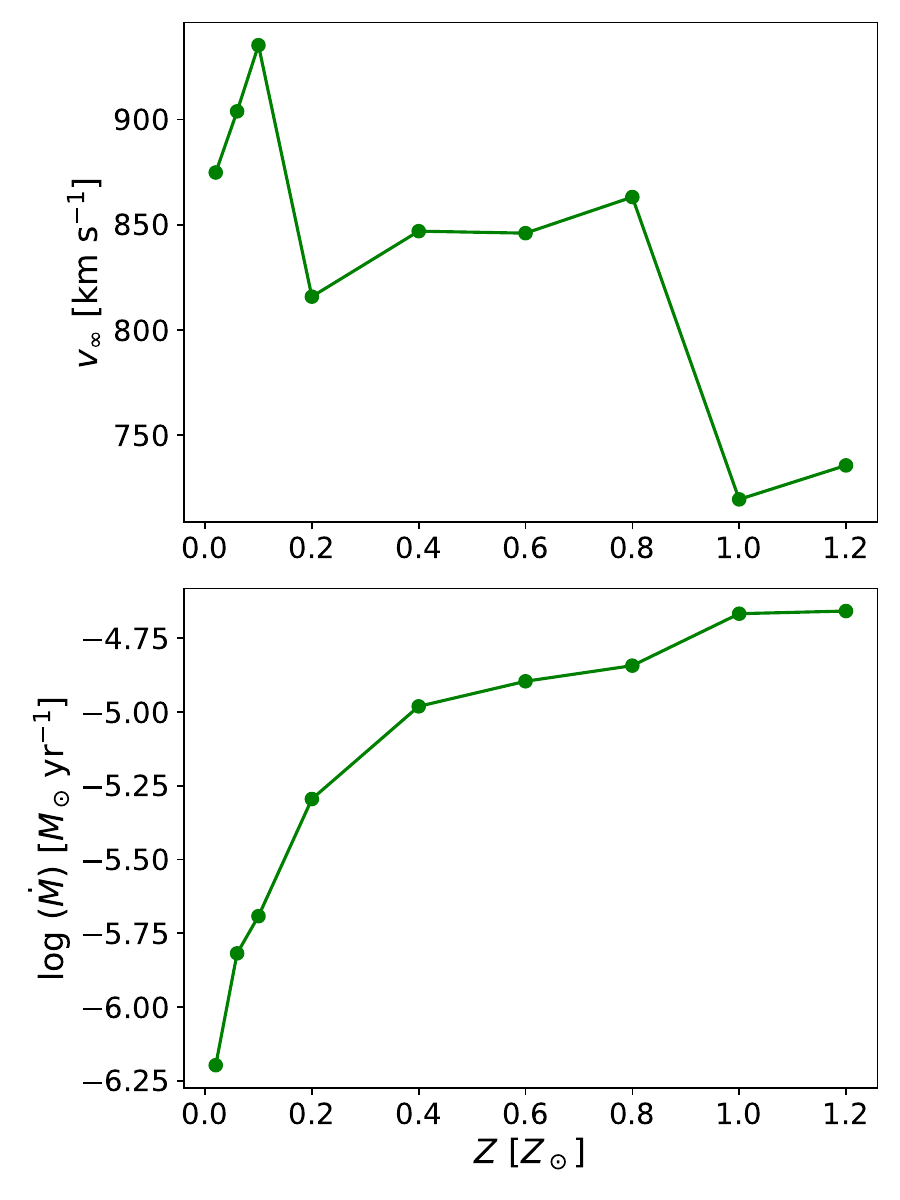}  
	\caption{Terminal wind velocities (top) and mass-loss rates (bottom) in terms of metallicity at $T_\ast = 26$ kK.}
	\label{fig:vfinals-zs-26kK}
\end{figure}

In Fig.\,\ref{fig:vfinals-zs-26kK}, we illustrate the $\dot{M}$- and corresponding $\varv_\infty(Z)$-trend resulting from the models with $T_\ast = 26\,$kK, which is a bit hotter than the aforementioned coolest models and thus not corresponding to the highest mass-loss regime at (very) low metallicity (cf.\,the bottom right panel in Fig.\,\ref{fig:cmp-mdtrends}). 
The overall behavior is still the same with $\varv_\infty$ tending to increase due to weaker $\dot{M}$, but also needs to decrease due to the reduction of available line opacity. 
Despite the huge change in $\dot{M}$, this example shows that $\varv_\infty$ mainly scatters around the same order of values despite the significant changes in $Z$. 
This approximately constant $\varv(Z)$ trend is not obtained for the hotter models, which behave more similar to the classical WR stars in \citet{sander-vink2020}. 
For example, the $35\,$kK models change from $\varv = 1209\,\mathrm{km\,s}^{-1}$ at $Z_\odot$ to $3408\,\mathrm{km\,s}^{-1}$ at $0.1\,Z_\odot$. 
This change by more than $2000\,\mathrm{km\,s}^{-1}$ is accompanied by a decrease in $\dot{M}$ by about 1.6\,dex, similar to what we see in Fig.\,\ref{fig:vfinals-zs-26kK} despite the very different $\varv$ behaviour.

\subsection{Comparison to Previous WNh Modelling}\label{subsec:comparison}

\begin{figure}
    \centering
    \includegraphics[width=\hsize]{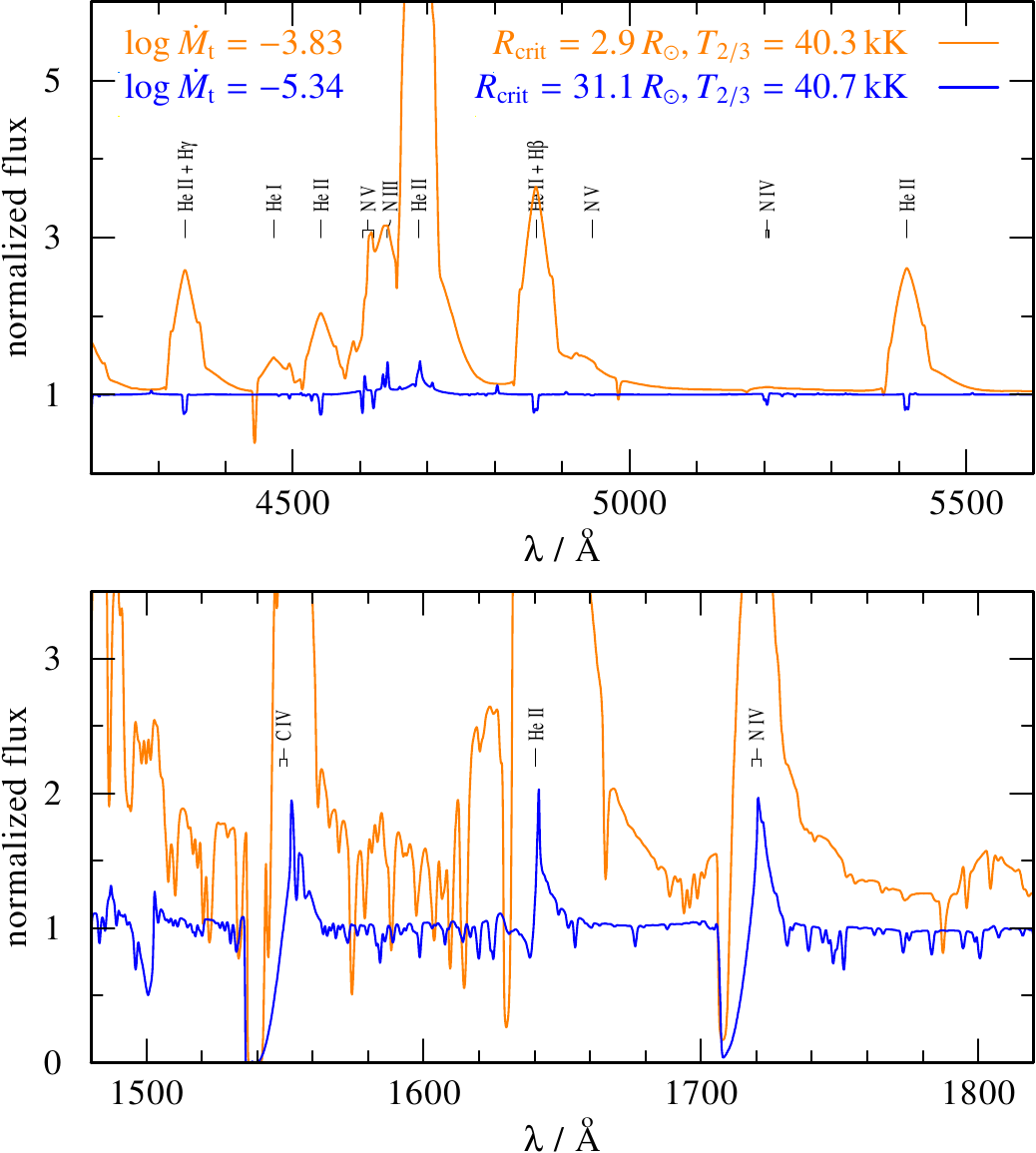}  
	\caption{Comparison of the optical (upper panel) and UV (lower panel) spectrum for a PoWR$^\textsc{hd}$ model from the current WNh sequence with shallow wind launching ($T_\ast = 42\,$kK, blue) compared to model with a similar $T_{2/3}$ (and $\varv_\infty$) but with deep wind launching ($T_\ast = 115\,$kK, orange). Both models share the same $L$, $M$, and chemical composition, but differ significantly in $\log \dot{M}_\text{t}$.}
	\label{fig:cmp-spec-deeplaunch}
\end{figure}

In Fig.\,\ref{fig:cmp-spec-deeplaunch}, we show an example from the current WNh sequence compared to a model where $\tau_\mathrm{max}=100$ and launches its wind from the hot iron opacity bump, but has otherwise the same fundamental stellar parameters and also a similar effective temperature $T_{2/3}$.
While the terminal velocities differ only by $\sim$$200\,\mathrm{km\,s}^{-1}$ ($2030\,\mathrm{km\,s}^{-1}$ for the deep model versus $2260\,\mathrm{km\,s}^{-1}$ for the shallow model), the mass-loss rates differ by $1.46\,$dex ($-3.80$ versus $-5.26$). 
The combination of $T_{2/3}$ and the line strengths in the spectrum can thus provide an immediate insight on the structure of the deeper atmosphere layers and the wind-launching regime. 
Hence, we can conclude that most known WNh stars residing on and below the main sequence cannot have winds launched by the hot iron bump.

\begin{figure}
    \centering
    \includegraphics[width=\hsize]{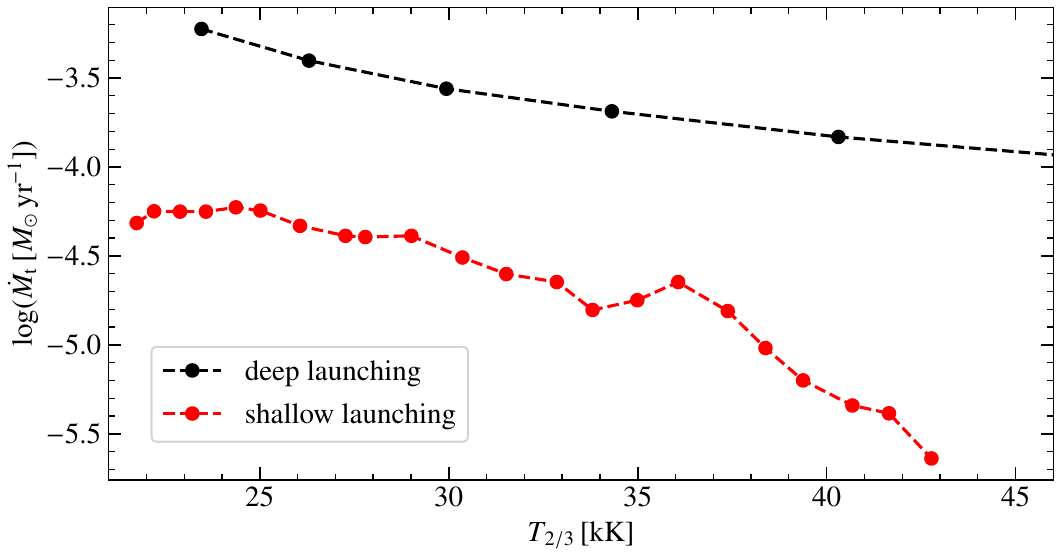}  
	\caption{Comparison of $\dot{M}_\text{t}$ versus $T_{2/3}$ between the $Z_\odot$ PoWR$^\textsc{hd}$ model sequence with shallow wind launching (red) and models with the same stellar parameters but deep wind launching (black).}
	\label{fig:cmp-mdtrends}
\end{figure}

The qualitatively different behavior of the wind regime studied in this work compared to the wind regime launched by the hot iron bump is also evident.
Shown in Fig.\,\ref{fig:cmp-mdtrends}, we compare the $T_{2/3}$-trend of the transformed mass-loss rate. 
For the comparison sequence at $Z_\odot$, we obtain a smooth downward trend with increasing $T_{2/3}$, while our model series shows a qualitatively different slope with a lot more substructure.

\begin{figure}
    \centering
    \includegraphics[width=\hsize]{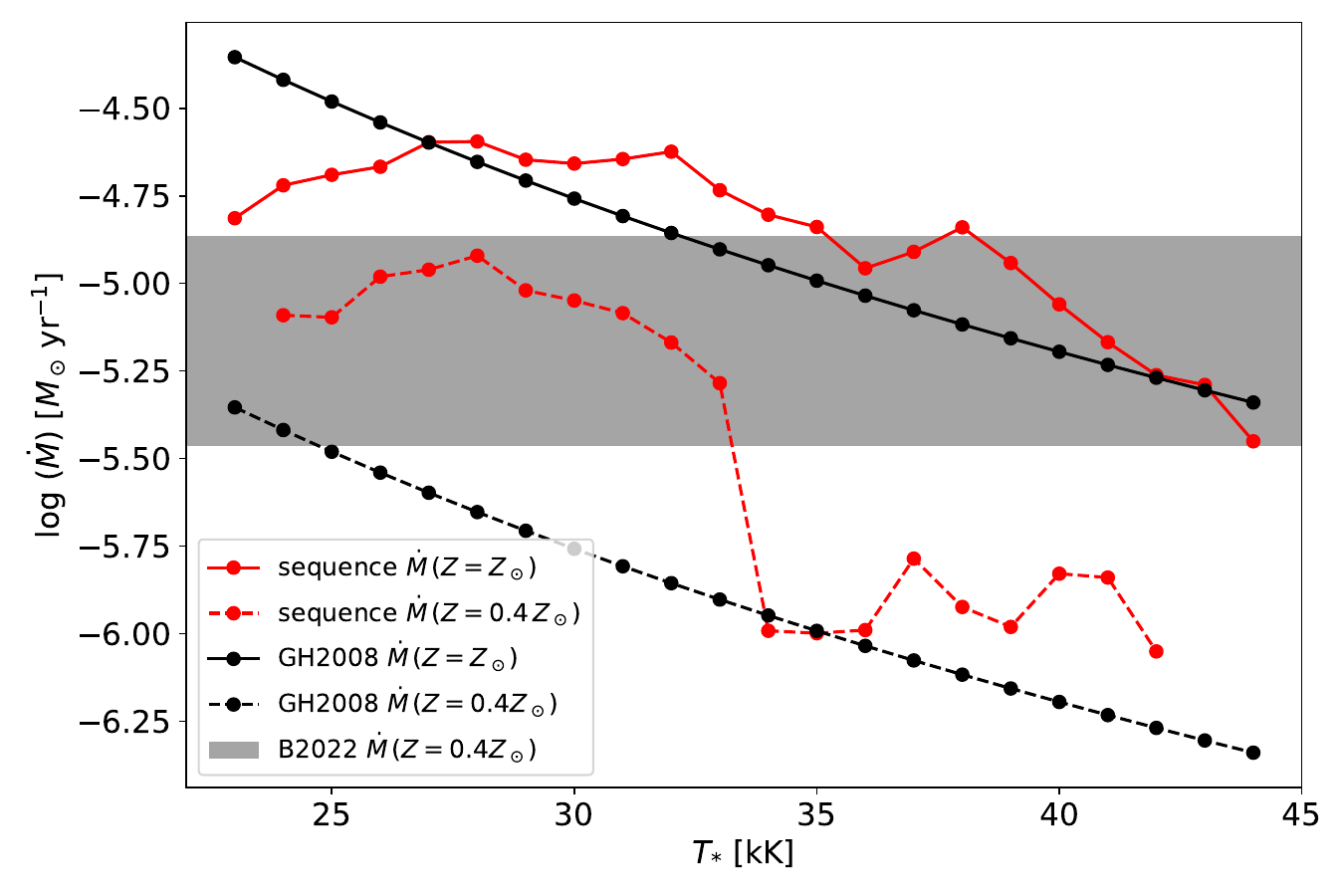}  
	\caption{Comparison of the $Z_\odot$ model sequence with the $\dot{M}(Z)$ recipe from \citet[GH2008]{graefener2008}, where we display the $Z=Z_\odot$ and $0.4\,Z_\odot$ cases (in black). 
    The shaded area represents the $\dot{M}$ range resulting from the recipe in \citet[B2022]{brands2022}, for the different $\Gamma_\mathrm{e}$ of our $Z=0.4$ sequence.}
	\label{fig:mdot-comp}
\end{figure}

Based on an set of dynamically-consistent calculated with an earlier version of PoWR (e.g., containing a more limited set of elements and ions), \citet{graefener2008} derived a mass-loss recipe for WNh-type stars. 
While their recipe is anchored on the analysis of the hotter WN7h star WR\,22 ($T_\ast = 44.7\,$kK), their models also cover a regime down to $31.6$\,kK. 
We extrapolate their derived formula and show the comparison of our models with their recipe in Fig.\,\ref{fig:mdot-comp} for two different metallicities. 
In the $Z = Z_\odot$ case, the $\dot{M}$ values of our model sequence are of similar order as their formula with the on average slightly higher values likely arising from the additional opacities included in the newer models.
While the formula from \citet{graefener2008} does reflect all the scatter in their individual models, there is actually more substructure in our model trends than in their model sets. 
While this partially might result from the additional complexity of the newer models, the prominent flattening of the $\dot{M}$-trend towards cooler temperatures does not occur in \citet{graefener2008} as this only occurs at the edge and beyond their grid coverage, i.e., for $T_\ast \leq 32\,$kK. 
This is even more impactful for the $0.4\,Z_\odot$-sequence, where the \citet{graefener2008} formula essentially only covers the regime of lower mass-loss rates. 
In fact, on the cooler side of the $\dot{M}$-jump we see almost an order of magnitude difference in the mass-loss rate for some temperatures.

More recently, \citet{bestenlehner2020} suggested a mass-loss recipe for the Of to WNh regime based on empirical results in the LMC Tarantula region. 
As there is no explicit metallicity and temperature dependence, we just highlight the covered $\dot{M}$ range in Fig.\,\ref{fig:mdot-comp}, using the calibration values from \citet{brands2022}. 
Notably, this region is in line with the cooler part of our $0.4\,Z_\odot$-sequence, i.e., the part showing higher mass-loss rates, while there is a difference of about half an order of magnitude on the hotter side. 
Yet, one has to consider that the nature of our modeled objects is likely different from most of the LMC R136 stars as those have higher hydrogen surface abundances, which would boost the resulting mass-loss rates (for the same $L/M$) as shown in \citet{sander2023,Sander2024iau}.

\subsection{Spectral appearance and WR definition}

\begin{figure}
    \centering
    \includegraphics[width=\hsize]{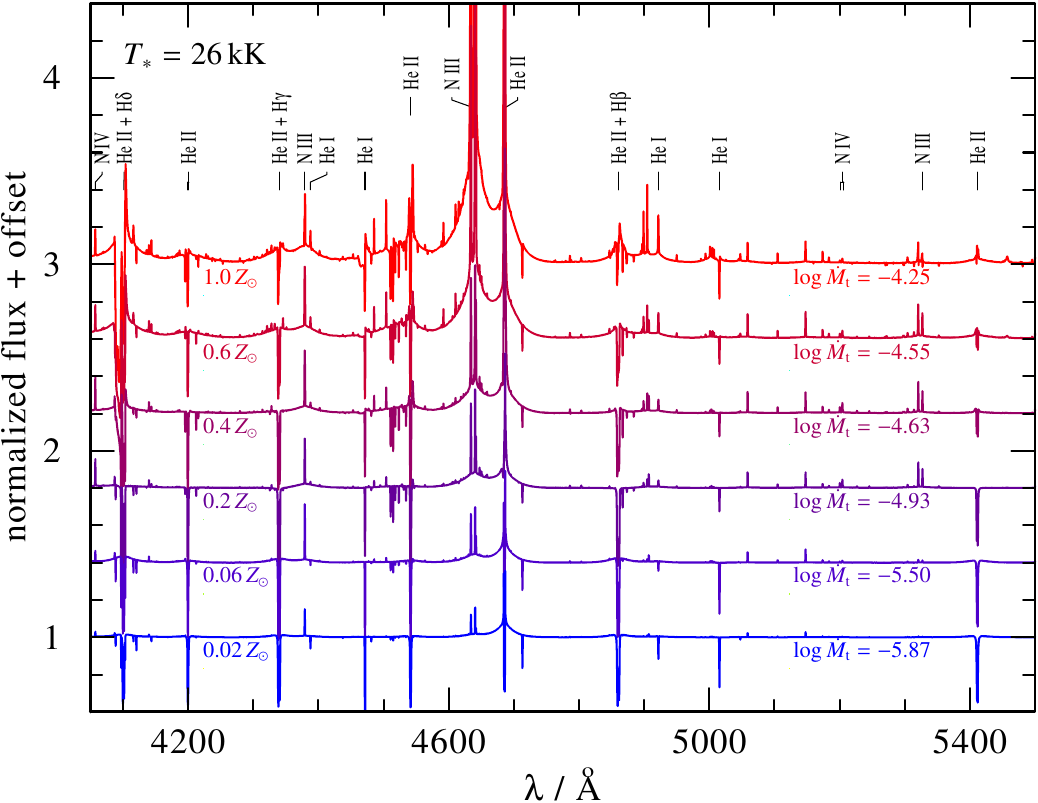}  
	\caption{Optical spectra of $T_\ast=26$ kK for different metallicities with offset by increasing metallicity.}
	\label{fig:cmp-spec-zseq}
\end{figure}

WR stars are formally a spectroscopic definition, which raises the question whether our models with low mass-loss rates actually qualify for being classified as WR. 
In fact, the criteria to be classified as WR vary between different temperature regimes. 
For the transition between early O and mid-type WNh stars, corresponding to objects located near or on the zero-age main sequence in the HRD, H$\beta$ is the main qualifier \citep{Crowther2011} with H$\beta$ in emission defining a WR star. 
Cooler objects with a mixture of absorption and emission lines are sometimes summarized under the Ofpe/WN9 label, but as \citet{crowther1995ofpe} and \citet{Crowther1997} argue one can naturally extend the WN sequence to lower ionization stages. 
As \ion{He}{ii}\,4686\,\AA\ can be in emission in O supergiants, they suggest using the occurrence of blue-shifted absorption or narrow emission features in \ion{He}{ii}\,4542\,\AA\ (or 5412\,\AA) and \ion{He}{i}\,4471\,\AA\ (or 5876\,\AA) to distinguish between wind-dominated and thus WN-defining spectra, and mainly photospheric spectra.

In Fig.\,\ref{fig:cmp-spec-zseq}, we show a sequence of spectra at different metallicities for the same $T_\ast$. 
At solar-like metallicities, the spectra clearly are clearly of WN type, but this changes with lower $Z$. 
At $0.2\,Z_\odot$, small emissions features remains for example in \ion{He}{ii}\,4542\,\AA, while \ion{He}{ii}\,5412\,\AA\ appears to be in pure absorption already. 
At $0.06\,Z_\odot$ and below, none of the lines except \ion{He}{ii}\,4686\,\AA\ is in emission, although some traces of electron scattering wings can still be noticed for a few \ion{He}{i} and \ion{He}{ii} lines given the proximity to the Eddington Limit. 
The spectra in Fig.\,\ref{fig:cmp-spec-zseq} correspond to the model sequence plotted in Fig.\,\ref{fig:vfinals-zs-26kK}, meaning that terminal velocities do not change much throughout this particular sequence.
These models, having late-type WNh-star characteristic parameters, no longer have spectra that will be seen as WR-star spectra at lower metallicities. 
From $Z\lesssim 0.2\,Z_\odot$ on, diagnostic lines in emission for WR stars are now in absorption, additionally signalling the winds are getting comparatively very thin in the low $Z$ regime. 
Hence, there is a transition from stars that show clear characteristics for late-type WNh stars at higher metallicities to stars with some wind imprint in the spectra but no longer a sufficiently strong wind to cause WR-like spectra.

%-----------------------------------------------------------------
\section{Summary and Conclusions}\label{sec:conclusions}

The in-depth analyses of late-type, hydrogen-rich WR star models with our $T_\ast$ sequences reveals a range of remarkable properties in this near-Eddington limit regime. 
Ranging from super-solar to sub-SMC metallicities, a myriad of effects are seen that do not all occur for other WR-star parameter ranges \citep[e.g. at higher $T_\ast$ in][]{sander2023}.
At solar metallicity, an overturn in both the mass-loss rates and wind efficiencies is seen; after an initial increase in $\dot{M}$ for rising $T_\ast$, the overall trend is decreasing for models where $T_\ast \gtrsim 25$ kK. 
In addition, we see similar behaviour for the wind efficiency, where the $\eta$-values peak at 32 kK instead.
This also occurs at super-solar metallicities, where the trends in global wind parameters are the same overall. 
For the sub-solar metallicity models, those with $0.4\,Z_\odot\leq Z \leq 0.8\,Z_\odot$, this same overturn for $\dot{M}$ and $\eta$ is observed, now coinciding at around the same $T_\ast$-values while the temperatures where these maxima occur drop systematically for decreasing $Z$. 
These maxima can be related to an increase in iron opacity $\varkappa_\mathrm{F}$ occurs due to \ion{Fe}{v} to \ion{Fe}{iv} recombination that tends to peak at the $T_\ast$ where the turnover occurs.
At SMC and sub-SMC metallcities $Z \lesssim 0.2\,Z_\odot$, such global wind maxima no longer seem to be present or fall below the temperature range of our models.

A striking feature we observe is the jump in $\dot{M}$ values for our $0.4\,Z_\odot\leq Z \leq 0.8\,Z_\odot$ model sequences.
Here, we can clearly distinguish our models in a `hot' and a `cool' side, where the wind characteristics differ significantly. 
On the cool side of these jumps, the winds typically show significantly higher $\dot{M}$ and lower $\varv_\infty$, vice versa on the hot side. 
The $T_\ast$ values where this jump occur range from 32-34 kK. 
This sharp difference in wind regime also translates to wind stratification (e.g. the velocity fields in Appendix \href{https://doi.org/10.5281/zenodo.15675082}{D} and the line accelerations in \href{https://doi.org/10.5281/zenodo.15675082}{C}) and spectral differences, the latter especially in the case of heavily wind-affected lines in the UV (see Appendix \href{https://doi.org/10.5281/zenodo.15675082}{B}). 
In a number of ways, this sharp transition of wind regimes is reminiscent of the classical bi-stability jump \citep[see e.g.][]{vink1999, vink2001}.
Indeed, studying the iron opacity $\varkappa_\mathrm{F}$ stratification in our wind models, we can clearly distinguish two opacity groups in the sequences where the jump occurs. 
However, while the classical bi-stability jump is described to be caused by \ion{Fe}{iv} recombining to \ion{Fe}{iii}, the situation in our models is of a more complex nature. 
While we generally do see a recombination of \ion{Fe}{v} to \ion{Fe}{iv} and a switch between lead iron ion driving in the temperature ranges of our model sequences, this does not necessarily lead to a sharp jump in wind behaviour, this latter being the case for $Z \geq Z_\odot$ models. 
Instead, we here distinguish two regimes on the basis of overall $\varkappa_\mathrm{F}$ considerations: on the cool side of the jump, these models typically have a clear $\varkappa_\mathrm{F}$ increase on and around the critical point $R_\mathrm{crit}$. 
On the hot side, this $\varkappa_\mathrm{F}$ tends to plateau instead.

These two main effects, the $\dot{M}$-$\eta$ turnover and the $\dot{M}$ jumps, do not occur clearly or at all for our lower-metallicity sequences at $Z \leq 0.2\,Z_\odot$.
At these $Z$, the contribution of iron to the wind driving starts diminishing and so do the effects of the iron opacities.
Here, other opacity sources take the upper hand in the driving of the wind; line opacities from e.g. \ion{N}{iii}, \ion{N}{iv} and \ion{C}{iv} along with continuum opacities from \ion{H}{I} and \ion{He}{II}. 
With the low impact of iron on the wind driving, the models start to adhere more to the results from studies of cWR stars \citep[e.g.,][]{sander2023}. 
At sub-SMC metallicities ($Z\leq 0.1\,Z_\odot$), the radiation driving does not suffice anymore to keep driving wind material further out in the wind. 
Instead, acceleration from the pressure gradient takes over as the main driver there.

These model sequences show significant differences to previous studies where the winds were typically launched deep, at high optical depths $\tau$. 
The main cause of this is that the models used in this study are shallowly launched, at $\tau$, in contrast to previous work where the winds are typically thought to be launched at the hot iron bump. 
Spectroscopically, our models tend to represent nature better with more physically realistic spectra, where there is a general order of magnitude drop in the mass-loss rates we find for this type of star. 
We note that our atmosphere models and the resulting spectra are representative for luminous, late-type WNh stars, thought to be core-H burning. 
Our results should not be directly extrapolated to all WNh-type stars as those could have very different combinations of stellar parameters causing a different wind behavior than we observe in our model sequences. 
Additional model sequences covering further parameter regimes will be required to combine our and other recent work into an improved mass-loss description for (cooler) WNh-type stars.

Given this range of effects and the insights on wind behaviour, along with the numerous spectral considerations from our analyses, future work warrants observational studies on late-type WNh stars that take our results into account. 
This is exactly the theme of a planned follow-up study, where previous results for WNh stars in the Galactic Centre will be compared in-depth with novel spectroscopic analysis  using hydrodynamically consistent modelling.

\section{Data Availability}

Appendices B (radiative accelerations), C (spectra) and D (velocity fields) and figures therein are found in \url{https://doi.org/10.5281/zenodo.15675082}.

\begin{acknowledgements}
AACS, MBP, and RRL are supported by the Deutsche Forschungsgemeinschaft (DFG, German Research Foundation) in the form of an Emmy Noether Research Group – Project-ID 445674056 (SA4064/1-1, PI Sander). AACS and RRL further acknowledge support from the Deutsche Forschungsgemeinschaft (DFG, German Research Foundation) – Project-ID 138713538
– SFB 881 (“The Milky Way System”, subproject P04). GNS and JSV are supported by STFC (Science and Technology Facilities Council) funding under grant number ST/Y001338/1. RRL is a member of the International Max Planck Research School for Astronomy and Cosmic Physics at the University of Heidelberg (IMPRS-HD).
NM acknowledges the support of the European Research Council (ERC) Horizon Europe grant under grant agreement number 101044048. 
F.N. acknowledges support by PID2022-137779OB-C41 funded by MCIN/AEI/10.13039/501100011033 by "ERDF A way of making Europe".
\end{acknowledgements}

%\nocite{*}
\bibliographystyle{aa}
\bibliography{paperbib.bib}

\begin{appendix}
% \FloatBarrier
\onecolumn
\section{Model abundances}%\label{app:specs}
In Table\,\ref{tab:abundances}, we provide an overview of the abundances and ions entering the atmosphere model calculations prestented in this work. The specific mass fractions listed in table\,\ref{tab:abundances} only refer to the sequence marked as solar metallicity $Z_\odot$.
For the sequences with different metallicities, we keep the same $X_\text{H}$, but scale metals (N to Fe) with $Z/Z_\odot$. The He mass fraction then follows as $X_\text{He} = 1 - X_\text{H} - \sum_{i=\text{N}\dots\text{Fe}} X_i$.

\begin{table}[h]
    \caption{Elements, ions and their respective abundances in mass fractions $X_\mathrm{m}$ used in this work for the $Z = Z_\odot$-sequence$^{a,b}$.}
    \centering
    % \flushleft
    \def\arraystretch{1.3}
    \label{tab:abundances}
    \begin{tabular}{lll}
        \hline
        Element & $X_\mathrm{m}$      & Ions \\
        \hline
        H       & (0.2)               & I, II \\
        He      & (0.63)              & I, II, III \\    
        N       & (0.015)             & I, II, III, IV, V \\   
        C       & ($1\cdot 10^{-4}$)         & I, II, III, IV, V, VI \\   
        O       & ($1\cdot 10^{-4}$)              & I, II, III, IV, V, VI \\
        Ne      & ($1.3\cdot 10^{-3}$) & I, II, III, IV, V, VI, VII \\
        Na      & ($2.7\cdot 10^{-5}$) & I, II, III, IV, V \\
        Mg      & ($7.0\cdot 10^{-4}$) & I, II, III, IV \\         
        Al      & ($5.7\cdot 10^{-5}$) & I, II, III, IV, V \\
        Si      & ($6.7\cdot 10^{-4}$) & I, II, III, IV, V, VI \\
        P       & ($5.8\cdot 10^{-6}$) & II, III, IV, V, VI \\
        S       & ($3.1\cdot 10^{-4}$) & I, II, III, IV, V, VI, VII \\
        Cl      & ($8.2\cdot 10^{-6}$) & III, IV, V, VI, VII \\
        Ar      & ($7.3\cdot 10^{-5}$) & I, II, III, IV, V, VI, VII, VIII \\
        K       & ($3.1\cdot 10^{-6}$) & I, II, III, IV, V, VI, VII \\
        Ca      & ($6.1\cdot 10^{-5}$) & II, III, IV, V, VI, VII, VIII \\
        Fe      & ($1.4\cdot 10^{-3}$) & II, III, IV, V, VI, VII \\
        \hline
    \end{tabular}
    \tablefoot{
        \tablefoottext{a}{Abundance values for CNO are adopted from the distribution typically found in WN stars and adopted in the public PoWR WN grids \citep[e.g.,][]{Todt2015}}
        \tablefoottext{b}{Other elements have been pulled from \citet{asplund2009} and \citet{Scott2015,Scott2015b} with Fe reflecting the ``iron group'' from Sc to Ni, summarizing their abundances.}
    }
    \label{tab:gridmodel_params}
\end{table}

\end{appendix}

\end{document}